\documentclass[aip,apl,a4,reprint,twoside,floatfix]{revtex4-1} 
\usepackage{graphicx}
\usepackage{amsmath}
\usepackage{amssymb}
\usepackage{amsfonts}
\usepackage{dcolumn}
\usepackage{epsfig}
\usepackage{subfigure}
\usepackage{bm}

\newcommand{\be}{\begin{equation}} \newcommand{\ee}{\end{equation}}
\newcommand{\bea}{\begin{eqnarray}} \newcommand{\eea}{\end{eqnarray}}

\begin{document}

\title {\bf Large structure-dependent room temperature exchange bias in self-assembled BiFeO$_3$ nanoparticles}

\author {Sudipta Goswami} \email{drsudiptagoswami@gmail.com} \affiliation {School of Materials Science and Nanotechnology, Jadavpur University, Kolkata 700032, India} 
\author {Aditi Sahoo} \affiliation {Advanced Mechanical and Materials Characterization Division, CSIR-Central Glass and Ceramic Research Institute, Kolkata 700032, India} 
\author {Dipten Bhattacharya} \affiliation {Advanced Mechanical and Materials Characterization Division, CSIR-Central Glass and Ceramic Research Institute, Kolkata 700032, India}
\author {Ozgur Karci} \altaffiliation{Current Address: The Institute of Optics, University of Rochester, New York 14620, USA} \affiliation{NanoMagnetics Instruments Limited, Suite 290, 266 Banbury Road, Oxford OX2 7DL, United Kingdom} 
\author {P.K. Mohanty} \affiliation {Department of Physical Sciences, IISER Kolkata, Mohanpur, West Bengal 741246, India} \affiliation {Saha Institute of Nuclear Physics, HBNI, 1/AF Salt Lake, Kolkata 700064, India}

\date{\today}

\begin{abstract}
We studied the magnetic properties of self-assembled aggregates of BiFeO$_3$ nanoparticles ($\sim$20-40 nm). The aggregates formed two different structures - one with limited and another with massive crosslinking - via `drying-mediated self-assembly' process following dispersion of the nanoparticles within different organic solvents. They exhibit large coercivity $H_C$ ($>$1000 Oe) and exchange bias field $H_E$ ($\sim$350-900 Oe) in comparison to what is observed in isolated nanoparticles ($H_C$ $\sim$ 250 Oe; $H_E$ $\sim$ 0). The $H_E$ turns out to be switching from negative to positive depending on the structure of the aggregates with $\mid +H_E \mid$ being larger. The magnetic force microscopy reveals the magnetic domains (extending across 7-10 nanoparticles) as well as the domain switching characteristics and corroborate the results of magnetic measurements. Numerical simulation of the `drying-mediated-self-assembly' process shows that the nanoparticle-solvent interaction plays an important role in forming the `nanoparticle aggregate structures' observed experimentally. Numerical simulation of the magnetic hysteresis loops, on the other hand, points out the importance of spin pinning at the surface of nanoparticles as a result of surface functionalization of the particles in different suspension media. Depending on the concentration of pinned spins at the surface pointing preferably along the easy-axis direction - from greater than 50\% to less than 50\% - $H_E$ switches from negative to positive. Quite aside from bulk sample and isolated nanoparticle, nanoparticle aggregates - resulting from surface functionalization - therefore, offer remarkable tunability of properties depending on structures. 
\end{abstract}

\maketitle

\section{Introduction}

The utility of self-assembled nanoparticle structures as against isolated nanoparticles has already been established in several cases \cite{Nie}. Most notable of them all are the plasmonic and magnetic properties which were utilized in optical and magnetic imaging devices employed in medical diagnosis and treatment. Research on utilization of self-assembled nanostructures in device applications has, so far, been focussed on metallic nanoparticles offering significantly different optical and magnetic properties from what is observed in isolated nanoparticles. For example, the plasmonic mode of Ag or Au nanoparticles exhibit either a blue or red shift depending on the self-assembled patterns formed through edge to edge or end to end connection \cite{Chen,Jain}. Likewise, the magnetic properties too exhibit pattern dependence via collective effects \cite{Lu,Tsymbal}. For instance, assembly of Co nanoparticles exhibit ferromagnetism with correlation length or domain size extending across 10 particles \cite{Alivisatos}. Depending on the areal density of the particles, the interaction among the particles could either be via exchange coupling or dipolar force. These observations triggered, in recent time, intense research on designing different self-assembled nanostructures for tailoring different physical properties. Apart from optical and magnetic properties, for example, very recently, role of self-assembled vertical stacking of 2D nanosheets of VOPO$_4$ and graphene has been shown \cite{Bando} to be extremely useful in allowing reversible intercalation of `beyond Li$^+$ ions' (such as Na$^+$, K$^+$, Al$^{3+}$, Zn$^{2+}$ etc) for developing zero-strain cathodes of batteries with enormous energy density and long cycle life. However, `structure dependence of properties' has still not been studied thoroughly for self-assembled structures of oxide nanoparticles (especially, perovskite ABO$_3$ type systems) which, either in bulk or isolated nanoscale form, exhibit a plethora of interesting electrical, magnetic, thermal, and optical properties \cite{Zubko}. In this paper, we report observation of remarkable variation in room temperature magnetic properties depending on nanostructures formed by aggregation of BiFeO$_3$ particles of size $\sim$20-40 nm. The results obtained for three different samples - isolated nanoparticles (Sample 1), nanoparticle aggregates with limited crosslinking (Sample 2) and massive crosslinking (Sample 3) - have been presented and discussed here.   

BiFeO$_3$ happens to be the most well-researched compound, either in bulk or nanoscale form, for its room temperature magnetoelectric multiferroic properties \cite{Catalan}. Bulk BiFeO$_3$ is a Type-I multiferroic with ferroelectric and magnetic transition temperatures $T_C$ $\sim$1103 K and $T_N$ $\sim$ 653 K. In a single crystal of BiFeO$_3$, spin flop transition could be noticed under sweeping electric field \cite{Lebeugle}. The epitaxial thin film, on the other hand, exhibits complete 180$^o$ switching of magnetic domains under sweeping electric field following a two-step process \cite{Heron}. The switching is associated with two-step rotation of ferroelectric domains - initially, by 71$^o$ and then by 109$^o$. The nanoparticles also exhibit significant change in ferroelectric polarization under a magnetic field \cite{Goswami-1}. However, following the observations made in self-assembled metallic nanoparticles, a simple question arises as to whether self-assembled aggregates of nanosized BiFeO$_3$ could exhibit different physical properties than what is observed in bulk, thin film, or isolated nanoparticles. Self-assembled nanocomposites of BiFeO$_3$-CoFe$_2$O$_4$, BiFeO$_3$-MgO or BiFeO$_3$-MgAl$_2$O$_4$ were synthesized earlier \cite{Comes,Stratulat,Aimon,Kim} by complicated techniques such as pulsed laser deposition combined with electron beam lithography. The spontaneous phase segregation between the two phases was used to design the self-assembled nanocomposites with enhanced magnetic and/or magnetoelectric properties. But, spontaneous formation of different nanoparticle aggregates of BiFeO$_3$ within different liquid media without the assistance of any template (template-free process) and structure-dependence of physical properties therein has not been studied so far. We report here that indeed room temperature magnetic properties depend significantly on the nanostructures formed within different liquid media via a process `drying mediated self assembly'. The nanoparticle aggregates exhibit sizable improvement in coercivity (by a factor of 4) with respect to what is observed in isolated particles and, more remarkably, a large spontaneous exchange bias which turns out to be switching from negative to positive depending on the structure. 

The exchange bias field has enormous applications in permanent magnet, magnetic recording media, anisotropic magnetoresistance, spin valves etc. Exchange bias field is observed in different spin composite systems containing ferromagnetic/ antiferromagnetic, ferrimagnetic/antiferromagnetic, ferromagnetic/spin glass interfaces \cite{Schuller}. It has also been observed in purely interface spin system \cite{Fan} where only the interface region - and not the bulk - exhibits finite exchange bias. During field cooling through the transition temperatures (ferromagnetic $T_C$ $>$ antiferromagnetic $T_N$), in conventional ferromagnetic/antiferromagnetic composites/heterostructures, unidirectional anisotropy of the moment develops at the interfaces because of exchange coupling between the moments across the interface. This unidirectional anisotropy, in turn, yields an exchange bias field \cite{Schuller}. It is reflected in the asymmetric shift of the magnetization versus field hysteresis loop along the field axis. Interestingly, apart from conventional exchange bias which requires field cooling through the transition temperatures ($T_C$ and $T_N$), exchange bias could be observed in many systems, over the years, even when the sample is cooled under zero field \cite{Keller,Ambrose,Wang,Coutrim}. This is called the $\textit{spontaneous exchange bias}$. The first field, applied for initiating the tracing of magnetic hysteresis loop, breaks the symmetry and creates the unidirectional anisotropy at the interface. It is similar to the hard-axis exchange bias where exchange bias is measured along `unmagnetized' hard axis following field cooling along the easy axis. The spontaneous exchange bias has been attributed \cite{Saha} to the energy landscape found in biaxial antiferromagnetic grains or in a pair of exchange-coupled uniaxial antiferromagnetic grains.

\section{Experimental}

The isolated nanosized particles were prepared by sonochemical route and subsequent heat treatment \cite{Goswami-2}. The x-ray diffraction data were recorded for the particles at room temperature. The particles were then dispersed within different organic solvents such as ethanol, iso-propyl alcohol, MEK+ethanol (azeotropic mixture), kerosene, and benzyl butyl phthalate in order to develop a variety of self-assembled patterns. The size of the particles varies within 20-40 nm while the concentration of the particles within the liquid medium was kept fixed at 10 mg/50 ml. The nanosized particles, depending on their size, shape, concentration within the liquid medium, and surface characteristics (surface charges, if any) in presence of ligand field of the surrounding liquid medium, undergo aggregation as a result of different interacting forces \cite{Bishop}. In the present case, it appears that the electrostatic, magnetic, and entropic forces play the major role in forming the aggregates. In order to examine whether the electrostatic force plays any role or not, we measured the zeta potential of the suspensions. The nanosized particles of BiFeO$_3$ develop positive surface charges due to the presence of Bi$^{3+}$ and Fe$^{3+}$ ions at the surface. The negative counterions from the surrounding liquid media form the double layer as testified by the observation of finite zeta potential. Interestingly, we observed that particles dispersed within ethanol (C$_2$H$_6$O) exhibit negative and very large zeta potential (-45.2 mV). On the other hand, for kerosene (C$_x$H$_y$; x = 6-16), the zeta potential, by crossing the isoelectric point, becomes positive and small (+15.7 mV). Finally, in the case of benzyl butyl phthalate (C$_{19}$H$_{20}$O$_4$), the zeta potential turns out to be still positive and even smaller (+6.8 mV). Based on this result, it is possible to conjecture that while ethanol forms a rather stable suspension with larger double layers and perhaps controlled aggregation (or even well-separated particles), in kerosene and benzyl butyl phthalate the size of the double layer drops and the nanoparticles aggregation is not quite well controlled. It turns out that the particles form structures with relatively limited crosslinking (Sample 2) within kerosene and massive crosslinking (Sample 3) within benzyl butyl phthalate which shows that longer the hydrocarbon chain of the liquid media smaller is the zeta potential and massive is the crosslinking in the aggregation of nanoparticles. Smaller zeta potential promotes generation of large number of nucleation centres for nucleation of different branches. It is important to mention here that the aggregated structures, in fact, consolidate and strengthen via drying of the liquid media. This is explained by the `drying-mediated self-assembly' mechanism \cite{Rabani}. The functional groups from the liquid medium attach with the surface of the nanoparticles. The liquid medium promotes mobility of the particles while progressive drying up of the liquid at different region of the entire suspension, thereafter, helps in forming the structures by consolidating the aggregation of the particles. Using transmission electron microscopy (TEM), the images of the nanostructures have been captured while high resolution TEM (HRTEM) was employed to observe the lattice fringes and their orientations. 

The room temperature magnetic hysteresis loops were measured by a Vibrating Sample Magnetometer (LakeShore Model 7407) while the magnetic force microscopy (MFM) was employed on the nanoparticles of BiFeO$_3$ to probe further the magnetic domain structure and its switching characteriscs under sweeping magnetic field across $\pm$20 kOe. The experiments have been carried out at the LT-AFM/MFM system \cite{Karci} of NanoMagnetics Instruments using commercial Co-alloy coated MFM cantilevers (PPP-MFMR, nominal coercivity of approximately 300 Oe \cite{Nanosensors}). The system uses two-pass mode of the MFM raster scan in which the cantilever is oscillated at the resonant frequency ($\sim$70 kHz) by a digital phase-locked-loop (PLL) control system at a certain oscillation amplitude of 10-50 nm. During forward scan, topography of the sample is recorded in the semi-contact mode using oscillation amplitude as a feedback parameter. At the end of the forward scan, the cantilever is lifted from the sample by about 10 to few hundred nanometers to get rid of the short-range forces. The phase shift of the cantilever, which is caused by the tip-sample magnetic interaction, is recorded as the magnetic image. The nanoparticles dispersed within suitable liquid media were deposited on glass substrate and dried subsequently.

\section{Theoretical Simulation}

To understand the underlying mechanism behind the nanoparticle aggregates formed in the present case, we consider a model of `drying-mediated-self-assembly' \cite{Rabani} on a $L\times L$ square lattice. Each cell  $i$ of the lattice  can be occupied by either liquid denoted by $l_i=1$ or vapour  $l_i=0.$ A  nanoparticle is considered to occupy $m \times m$ sites of the lattice, and its presence and absence at a site $i$ is denoted by $n_i = 1, 0$ respectively.  The Hamiltonian that captures the interactions among particles is given by \cite{Rabani}
\begin{equation}
 H=-\epsilon_l \sum_{\langle ij\rangle} l_i l_j 
 -\epsilon_n \sum_{\langle ij\rangle} n_i n_j
 -\epsilon_{nl}\sum_{\langle ij\rangle} n_i l_j - \mu \sum_i l_i
\end{equation}
where the first (second) term corresponds to the interaction among the neighboring liquid (nanoparticles) and the third term represent particle-liquid  interaction. Evaporation of the liquid is controlled by the potential $\mu$. We follow the dynamics of the system using Monte Carlo simulations where three basic steps that change the configurations of the system (namely diffusion of nano particles, diffusion of liquid and evaporation of liquid) are carried out stochastically with probability $p = min[1, exp(-\Delta H/k_B T)]$, where $\Delta H$ is the change in energy and $k_B$ is the Boltzmann constant. On a $500 \times 500$ square lattice, we consider each nanoparticle to occupy $3 \times 3$ cell which correspond to the size $\sim$20 nm, as observed experimentally. The coverage of this two-dimensional lattice was kept fixed at 20\% which, again, simulates the concentration 10 mg/50 ml used in the experiments. The interaction potentials ($\epsilon$) were considered to be $\epsilon_l = 1, \epsilon_n = 1,$ and $\mu  = -2.252$  while the particle-liquid interaction potential $\epsilon_{nl}$ was varied; it turns out that $\epsilon_{nl} = 0.1, 1,$ and $4$ simulate the attractive potential generated from the electrostatic, magnetic, and entropic interactions in different media such as ethanol, kerosene, and benzyl butyl phthalate. Starting from an initial configuration at time $t = 0$, where all the cells are occupied by liquid and 
20\% of the cells are filled by nanoparticles placed randomly, we monitor  the system for $t$ = 1000 MCS at $k_B$T = 0.2. The process dynamics is governed by  
the competition of two time scales - the evaporation and particle diffusion time
scales. Evaporation of the liquid in the adjacent cells prevents further particle diffusion and eventually forces nanoparticles to form a static pattern. The detailed structure of the patterns depends crucially on the inter-particle  interaction strength $\epsilon_{nl}$. In contrast to Ref. [26], we have used, in our case, biased diffusion of nanoparticles in a preferred direction which engineers the formation of crosslinking.

Interestingly, the simulated patterns could nearly replicate the experimental patterns. This shows that the particle-liquid interaction potential $\epsilon_{nl}$ indeed increases substantially as the suspension medium is changed from ethanol to kerosene, and then to  benzyl butyl phthalate; other interaction parameters $\epsilon_{l}, \epsilon_{n}$ and $\mu$ remain constant (or change only a little). Of course, direct experiments, apart from the measurement of zeta potential, is desirable in order to determine the overall interaction potential $\epsilon_{nl}$. Patterns observed from the simulations, though suitable only for qualitative comparison with the experimental results, still provide valuable insight about the trend of variation of $\epsilon_{nl}$ as the medium of suspension is changed.

 \begin{figure}[ht!]
 \begin{center}
 	\includegraphics[scale=0.20]{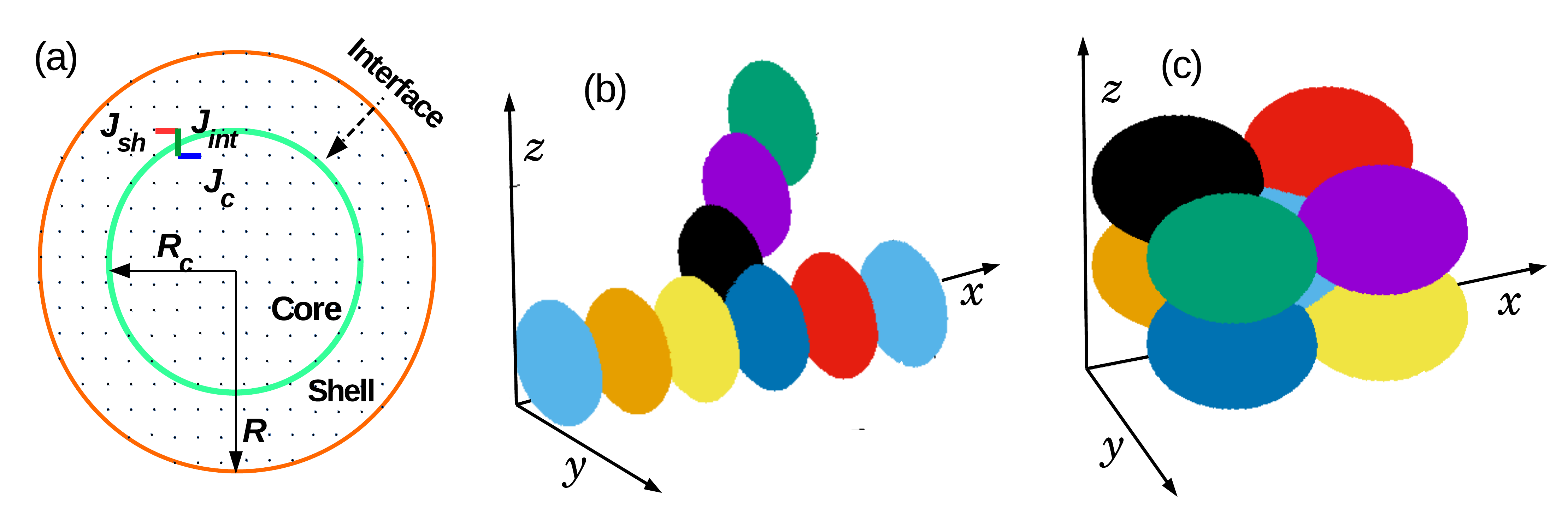}
 	\caption{(a): Cross-sectional view of a spherical nanoparticle of radius $R$ having a circular core of radius $R_c<R$. The annular region of width  $R-R_c$ represents the shell. Heisenberg spins within the core(shell) interact with strength $J_c$ ($J_{sh})$ and the interaction strength across the interface is $J_{int}$. A fraction of spins on the outermost surface (blue line) is pinned. (b) A crosslinked assembly of $N=9$ nanoparticles. (c) A closed packed or a massive crosslinked assembly.}
 	\label{fig:Fig1}
 	\end{center}
\end{figure}

As we have seen above, in drying conditions, the nanoparticles  form  different kind of aggregate structures depending on the solvent: isolated when the solvent is ethanol, linear structures with little or no crosslinks when the solvent is kerosene and a massive crosslinked nanostructure when the solvent is benzyl butyl phthalate (BBP). It has already been observed \cite{Berkowitz} that the spins on the surface of magnetic compounds in nanoscale, mainly in ferrite nanostructures, can get pinned in presence of organic solvents. Besides the structural organization, the surface pinning too may play a significant role in modifying the magnetic properties of nanoparticles. Keeping these facts in  mind, we modeled the inverse core-shell nanoparticles and their assembly in three dimension (3D). The effect of surface  pinning on nanostructures, including heterostructures, has been thoroughly discussed in a recent article \cite{Sahoo}. It was found  that the change in the magnetic properties due to surface pinning is  quite robust, in a sense that it is effective in  both two- and three-dimensional particles, and with different spin interactions such as Ising, XY, and Heisenberg. We consider spherical nanoparticles of radius $R$, having a spherical core ${\bf C}$ of radius $R_c<R$ as shown in  Fig. 1(a). The shell ${\bf Sh}$ of the nanoparticle is the annular region of width $R-R_c$. We assume that inside the nanoparticle, both in the core and the shell, classical Heisenberg spins  ${\bf S}_i$ are arranged in a cubic lattice, i.e., $2R+1$ spins along the diameter. Further, we consider that $N$ such identical nanoparticles form a superstructure, either as crosslinked linear chains (shown in Fig. 1(b)) or as closed packed structures (as  in Fig. 1(c)). To have a fair comparison of the  magnetic properties we have taken same number of nanoparticles, in total  $N=9$, and formed two different structures.  

Each lattice site $i$ of the supperstructure is associated with a classical Heisenberg spin ${\bf S}_i$ and the spins interact following the Hamiltonian \cite{Sahoo},
\bea
 {\cal H} &=&   -J_c \sum_{i\in {\bf C}, j\in {\bf C}} {\bf S}_i. {\bf S}_j  -J_{sh} \sum_{i\in {\bf Sh}, j \in {\bf Sh}}  {\bf S}_i. {\bf S}_j \cr &&-  J_{int}\sum_{i \in {\bf Sh}  j \in {\bf C}} {\bf S}_i. {\bf S}_j   - H \sum_{i\in {\bf C}, i\in {\bf Sh} } S_i^z,
 \label{eq:H}
 \eea
where $j$ is the nearest neighbor of site $i,$;  $J_c$, ($J_{sh}$) are the exchange interaction energies among the spins within the core (shell); $J_{int}$   represents  core-shell interface interaction, and $H$ is the external magnetic field. We aim at studying an inverse core-shell magnetic nanostructure, which can be modeled by taking  $J_c<0$  (antiferromagnetic core) and $J_{sh}>0$ (ferromagnetic shell). Pinning on the surface, occurring from the interaction of surface spins with solvents, can be modeled by two parameters, (i) $\eta$, representing a fraction of the surface spins which are frozen (i.e., they do not evolve  with time) and (ii) $r$  which denote the fraction of these frozen spins oriented preferenially along the positive easy-axis direction. The magnetic field is applied along the easy-axis, which is chosen as the $z$-direction. 

We study the hysteresis properties of these nanoparticle aggregate structures using Monte Carlo simulations with a single spin Metropolis algorithm where a trial configuration is accepted  with probability  $ Min\{ 1, e^{-\beta \Delta E}\}$. Here, $\Delta E$ is energy difference between the present configuration and the trial one. The trial configuration is constructed by changing the angles of a randomly chosen spin by a small but random amount. 

To calculate the hysteresis, first we set the temperature as $\beta^{-1}=1$, which is much smaller than the critical temperature $T_C$. This can be achieved from any  random initial configuration by relaxing the system in zero-field-cooled condition for a long time. The magnetic field is then raised slowly from $H=0$ to $H_{max}$ with a field sweep rate $\Delta H$ units per Monte Carlo sweep (MCS) and finally, the hysteresis loop calculations are undertaken for a cycle by varying the field from $H_{max}$ to $ -H_{max}$ and then to $ H_{max}$. Magnetization of the system is measured after each MCS and it is averaged over 100 samples.

\begin{figure}[ht!]
\centering
\includegraphics[scale=0.30]{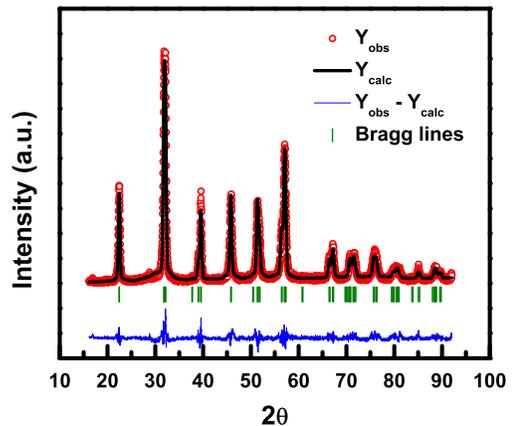}
\caption{The room temperature x-ray diffraction pattern for the isolated nano-sized particles of BiFeO$_3$.}
\end{figure}

\begin{figure*}[ht!]
\begin{center}
   \subfigure[]{\includegraphics[scale=0.10]{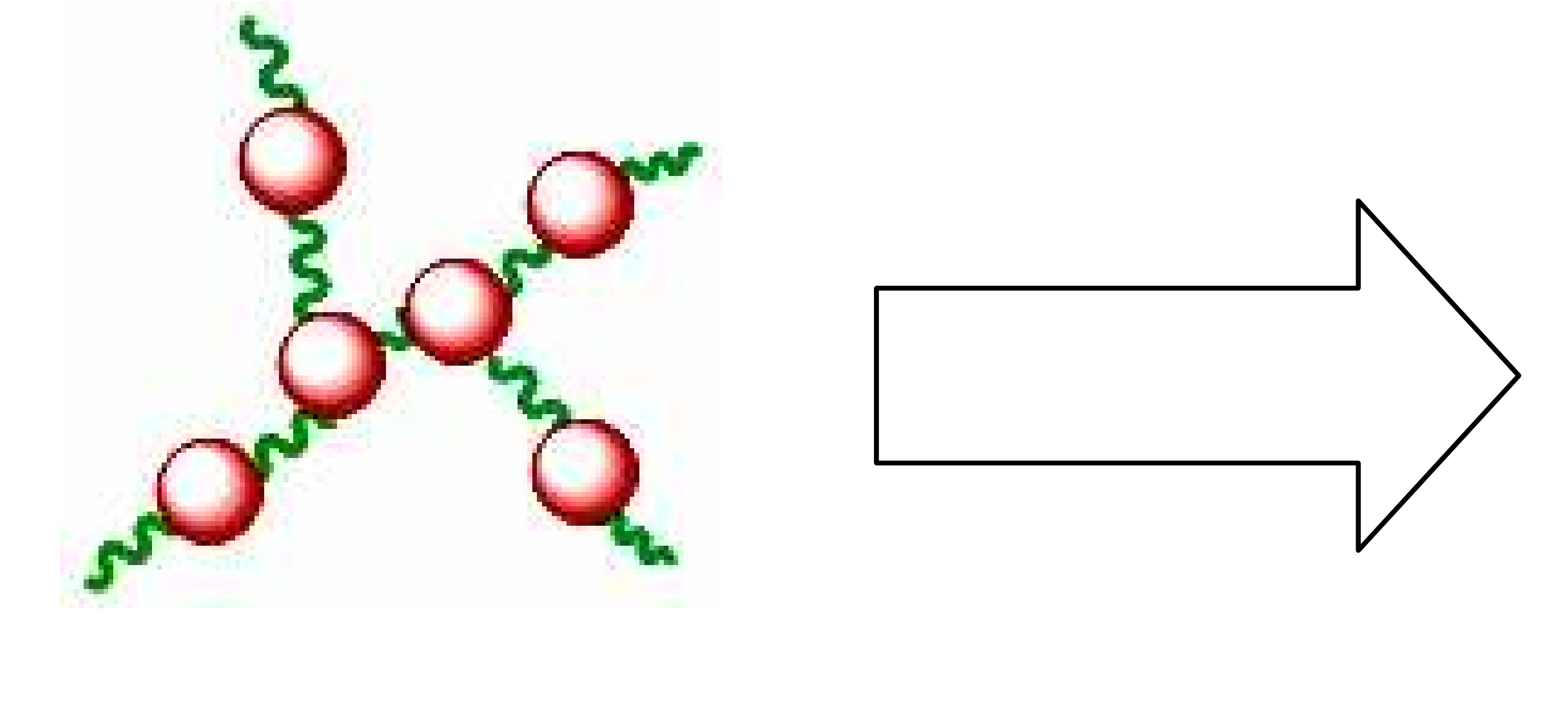}} 
   \subfigure[]{\includegraphics[scale=0.20]{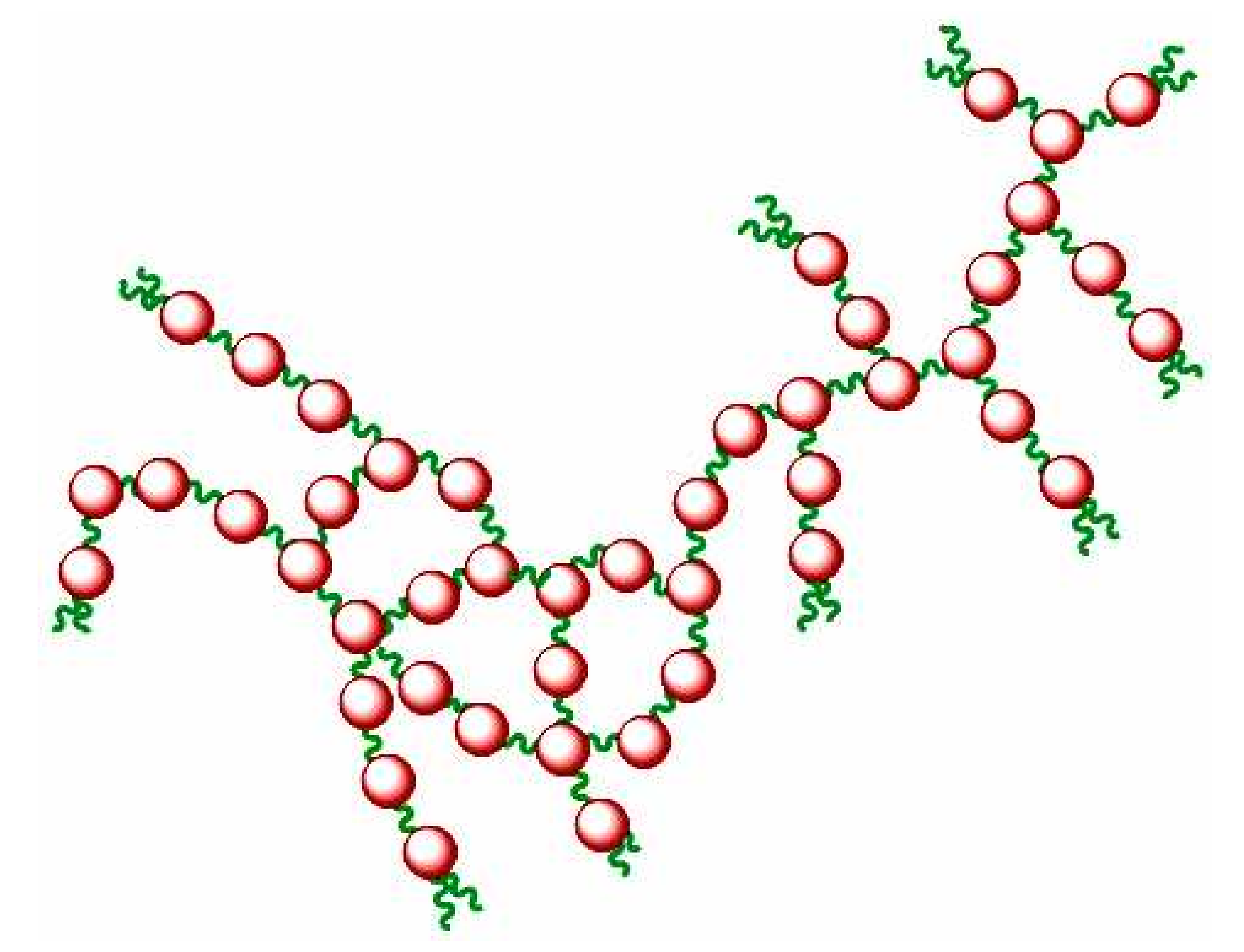}}
   \subfigure[]{\includegraphics[scale=0.10]{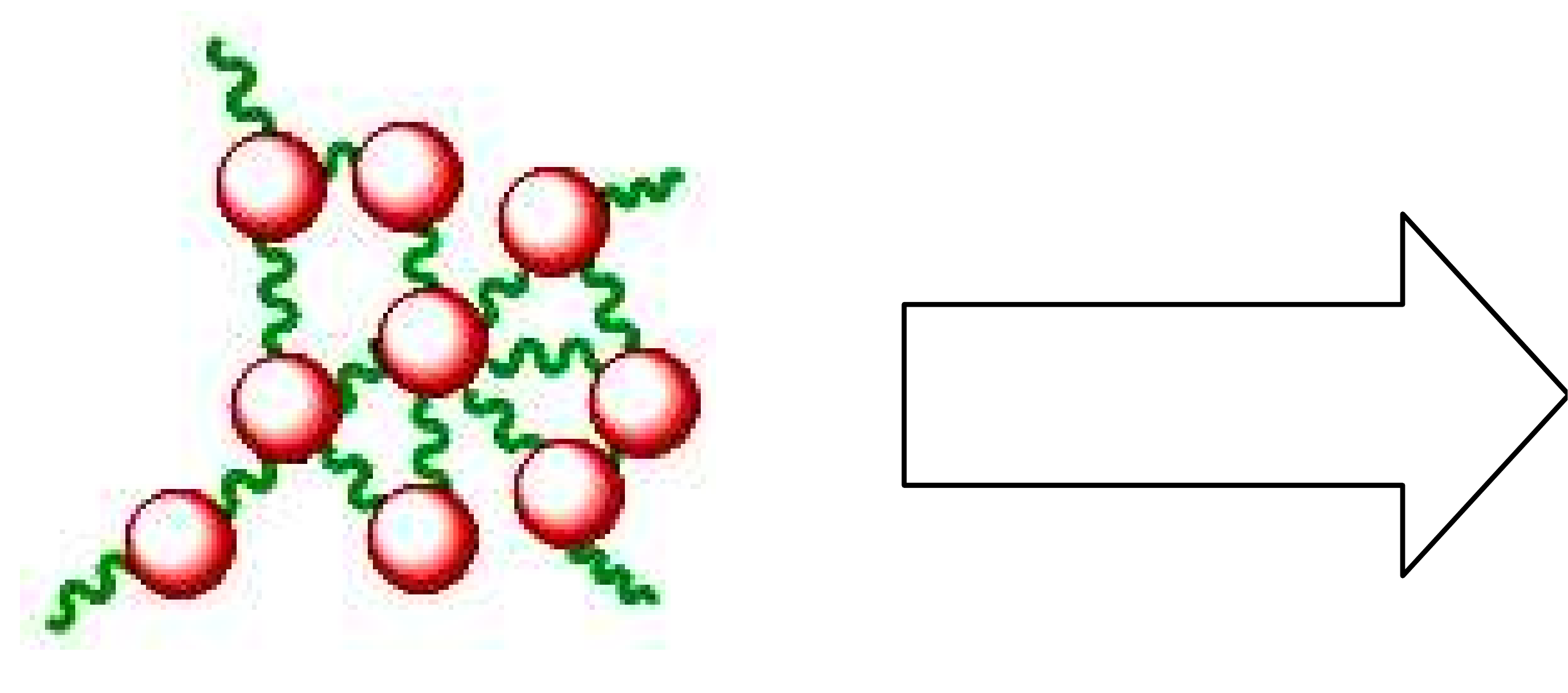}} 
   \subfigure[]{\includegraphics[scale=0.15]{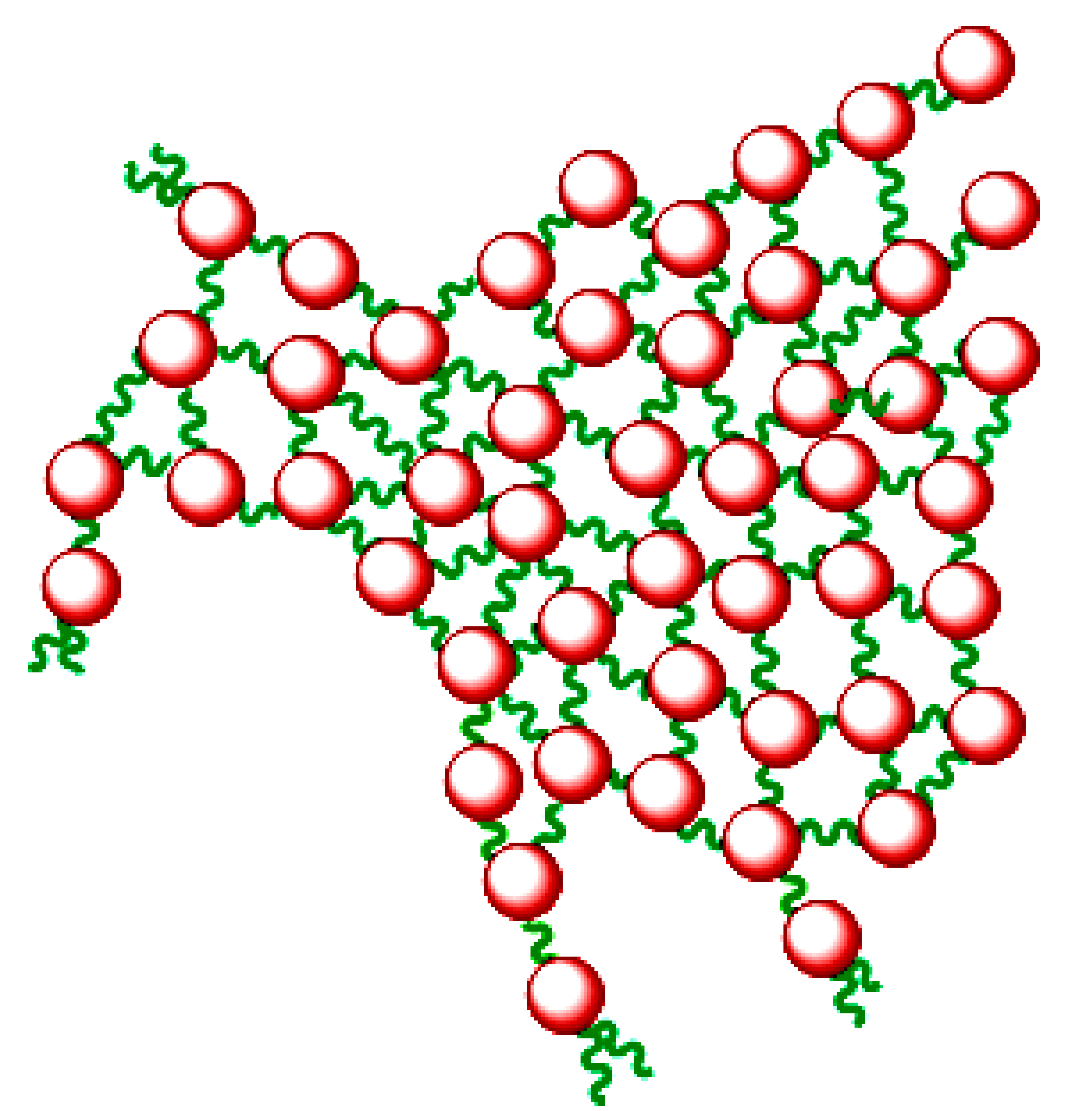}}
   \end{center}
\caption{The schematic illustration of the formation of different structures for BiFeO$_3$ nanoparticle aggregates; (a) shows the nucleation of the structure with limited crosslinking which eventually forms the structure (Sample 2) shown in (b); (c) shows the nucleation of the structure with massive crosslinking which eventually forms the structure (Sample 3) shown in (d); the nanoparticle aggregates with limited and massive crosslinked structure, as shown in (c) and (d), form within kerosene and benzyl butyl phthalate, respectively. }
\end{figure*}

\begin{figure*}[ht!]
\begin{center}
   \subfigure[]{\includegraphics[scale=0.45]{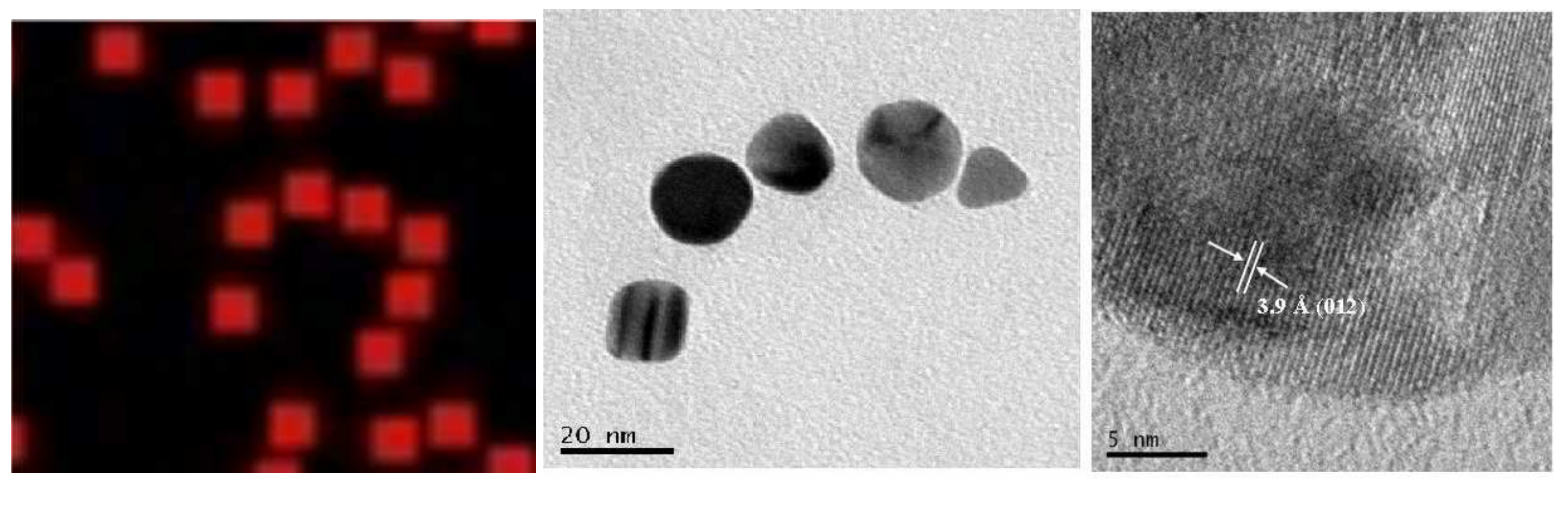}} 
   \subfigure[]{\includegraphics[scale=0.45]{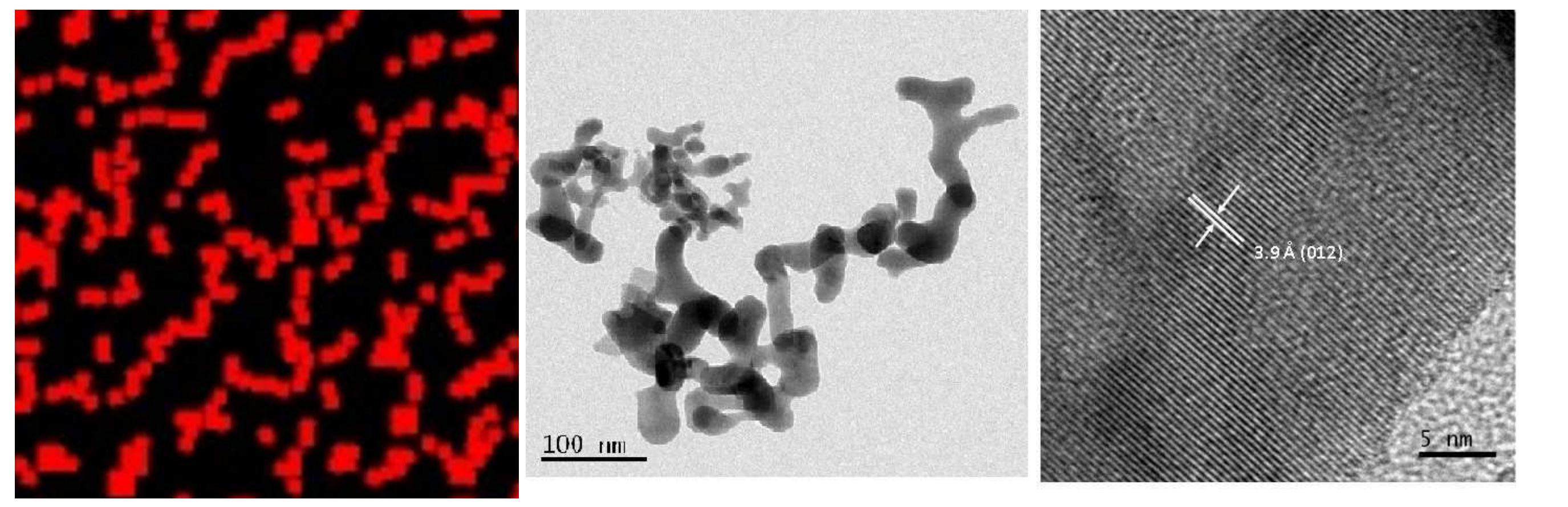}}
   \subfigure[]{\includegraphics[scale=0.45]{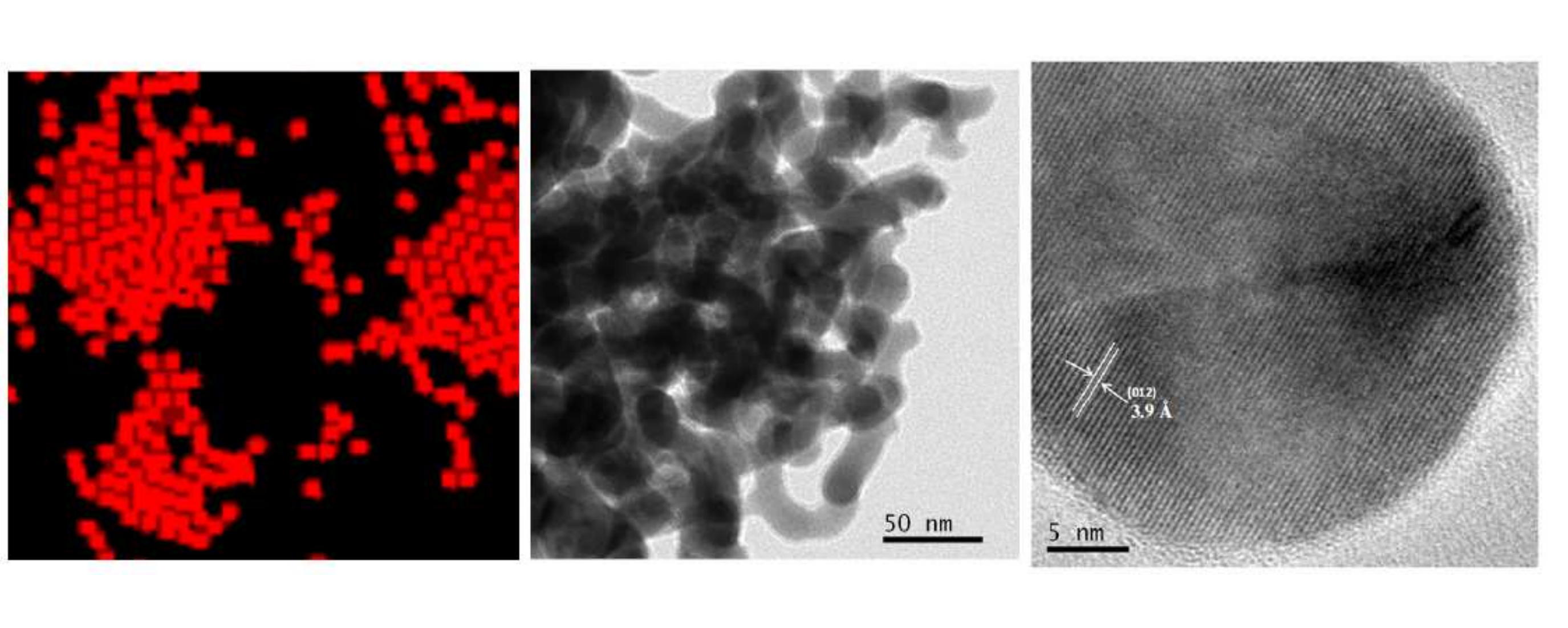}}
   \end{center}
\caption{The panels in (a) show the simulated structure for Sample 1 (left), its experimental bright field TEM image (center), and experimental HRTEM image with lattice fringes of (012) plane (right); panels in (b) show the simulated structure for Sample 2 (left), its experimental bright field TEM image (center), and experimental HRTEM image with lattice fringes of (012) plane (right); panels in (c) show the simulated structure for Sample 3 (left), its experimental bright field TEM image (center), and experimental HRTEM image with lattice fringes of (012) plane (right). }
\end{figure*}

\section{Results and Discussion}

Figure 2 shows the x-ray diffraction pattern of the isolated nanoparticles of BiFeO$_3$. They are phase pure (rhombohedral, space group R3c which supports both the polar and magnetic orders). They are then dispersed in different liquid media as discussed in the experimental section. The general features of the nanostructures are formation of nanoparticle aggregates with either limited or massive branching (crosslinking). The schematic illustration of the nanoparticle aggregate formation within different organic solvents is shown in Fig. 3. The aggregate with `limited crosslinked' structure (Sample 2) forms within kerosene via nucleation of limited number of branching (Fig. 3a) and their growth to form the structure is shown in Fig. 3b. Within benzyl butyl phthalate, on the other hand, nucleation of large number of branching (Fig. 3c) leads to the formation of `massive crosslinked' structure (Sample 3) (Fig. 3d). The particles eventually join to form those structures following drying of the liquid medium. Figure 4 shows the typical TEM and HRTEM images along with the simulated structures for Sample 1 (Fig. 4a), Sample 2 (Fig. 4b), and Sample 3 (Fig. 4c). In each case, while left panels show the structures simulated theoretically, the center and right panels, respectively, show the actual experimental TEM and HRTEM images. More examples are available in the supplementary material. The qualitative similarity between the simulated and experimental patterns is noteworthy. The simulated patterns across the full size 500 $\times$ 500 cells (each cell size $\xi$ = 7 nm) are shown in the supplementary material. The movies showing the self-assembly process over a time scale $\tau_s$ = 20s are also included as videos for all the three samples. As pointed out earlier, an important systematic in the pattern formation is the dependence on the length of C-H chain. Longer the C-H chain, massive is the crosslinking of the nanoparticles in the assembly. The particle-liquid interaction potential $\epsilon_{nl}$ enhances systematically with the enhancement of C-H chain length even when other potentials are kept constant. This, in fact, drives the formation of different nanostructures (Sample 2 and Sample 3) since smaller potential yields structures with limited crosslinking (Sample 2) while larger potential results in massively crosslinked structure (Sample 3) due to nucleation of particle aggregation or chain formation from many different nucleation sites. The high resolution TEM (HRTEM) experiments were also carried out (Fig. 4) for all the particles - either isolated or aggragated. The isolated particles are found to be oriented with either (012) or (110) planes perpendicular to the beam. Depending on the shape of the particles from nearly spherical to irregular, the lattice strain varies within 0.5\%-3.0\%. In an aggregated structure too, the particles appear to remain oriented with either (012) or (110) planes perpendicular to the beam. Therefore, collective effect in an aggregation has not induced any reorientation of the particles. 

\begin{figure}[ht!]
\begin{center}
   \subfigure[]{\includegraphics[scale=0.20]{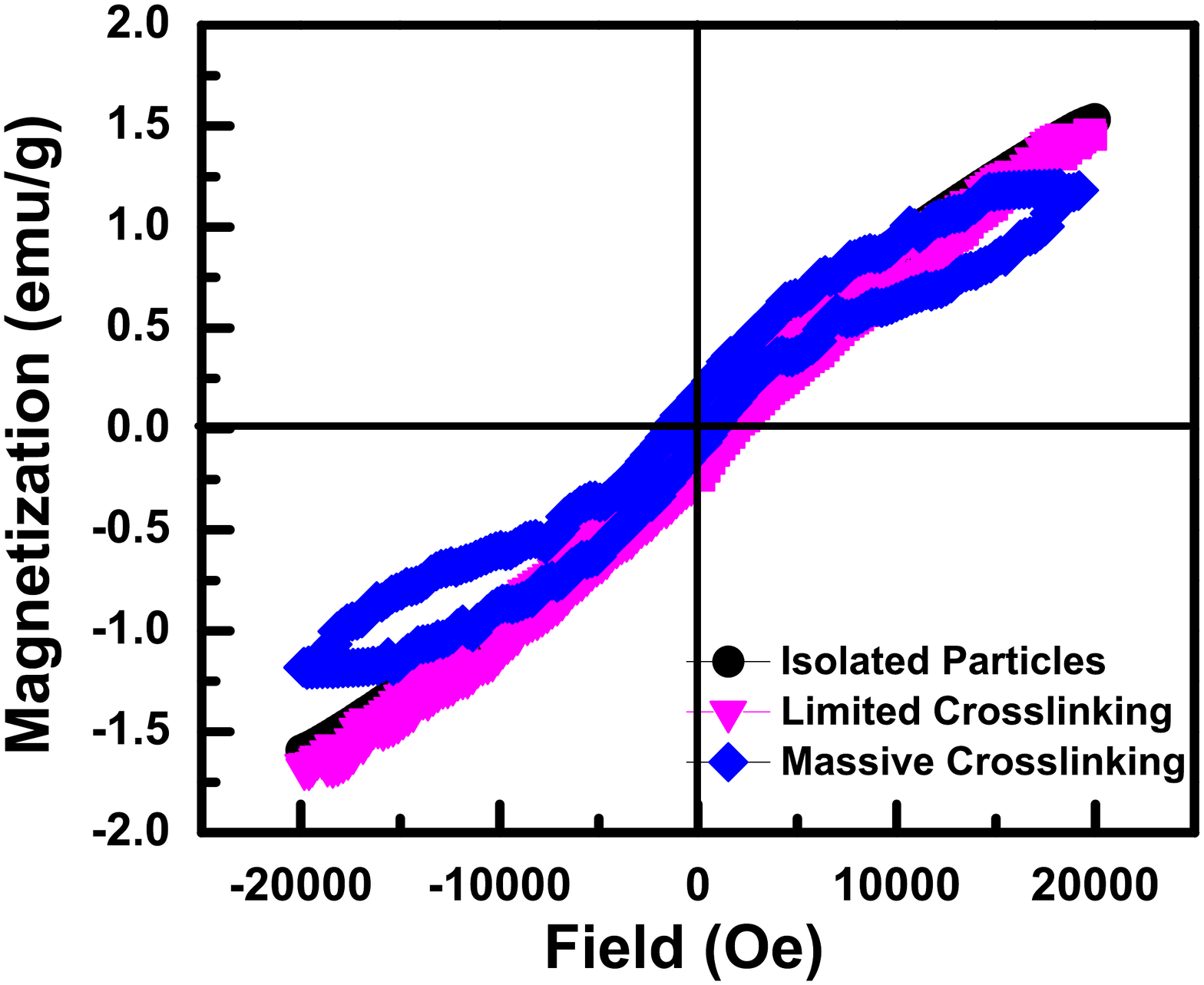}} 
   \subfigure[]{\includegraphics[scale=0.20]{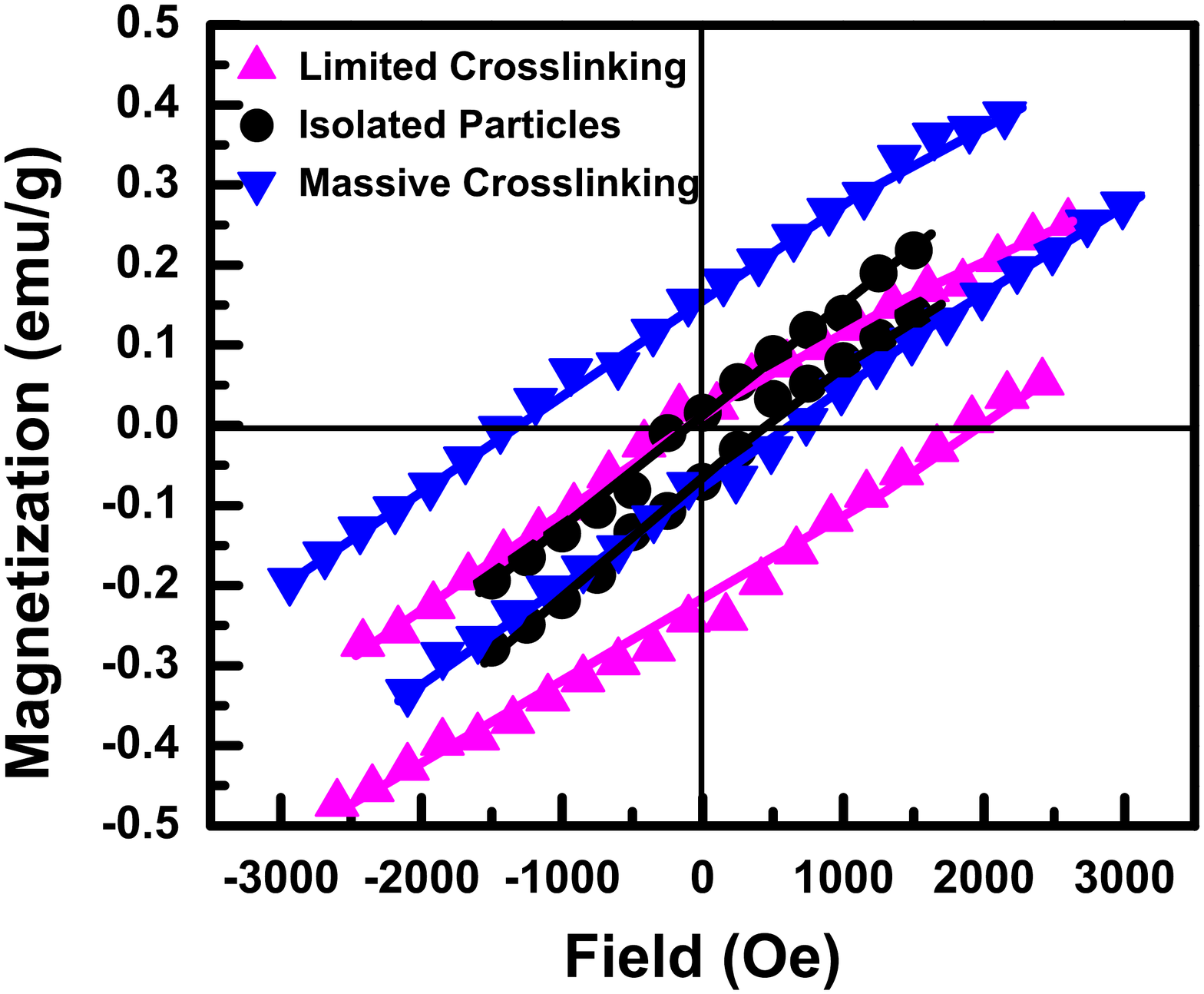}}
   \end{center}
\caption{(a) The magnetic hysteresis loops measured at room temperature on isolated nanosized particles (Sample 1) as well as on nanoparticle aggregates with limited and massive crosslinked structures (Sample 2 and Sample 3); (b) the low field region is blown up to show the coercive fields and exchange bias clearly.}
\end{figure}

\begin{figure}[htp!]
\begin{center}
   \subfigure[]{\includegraphics[scale=0.25]{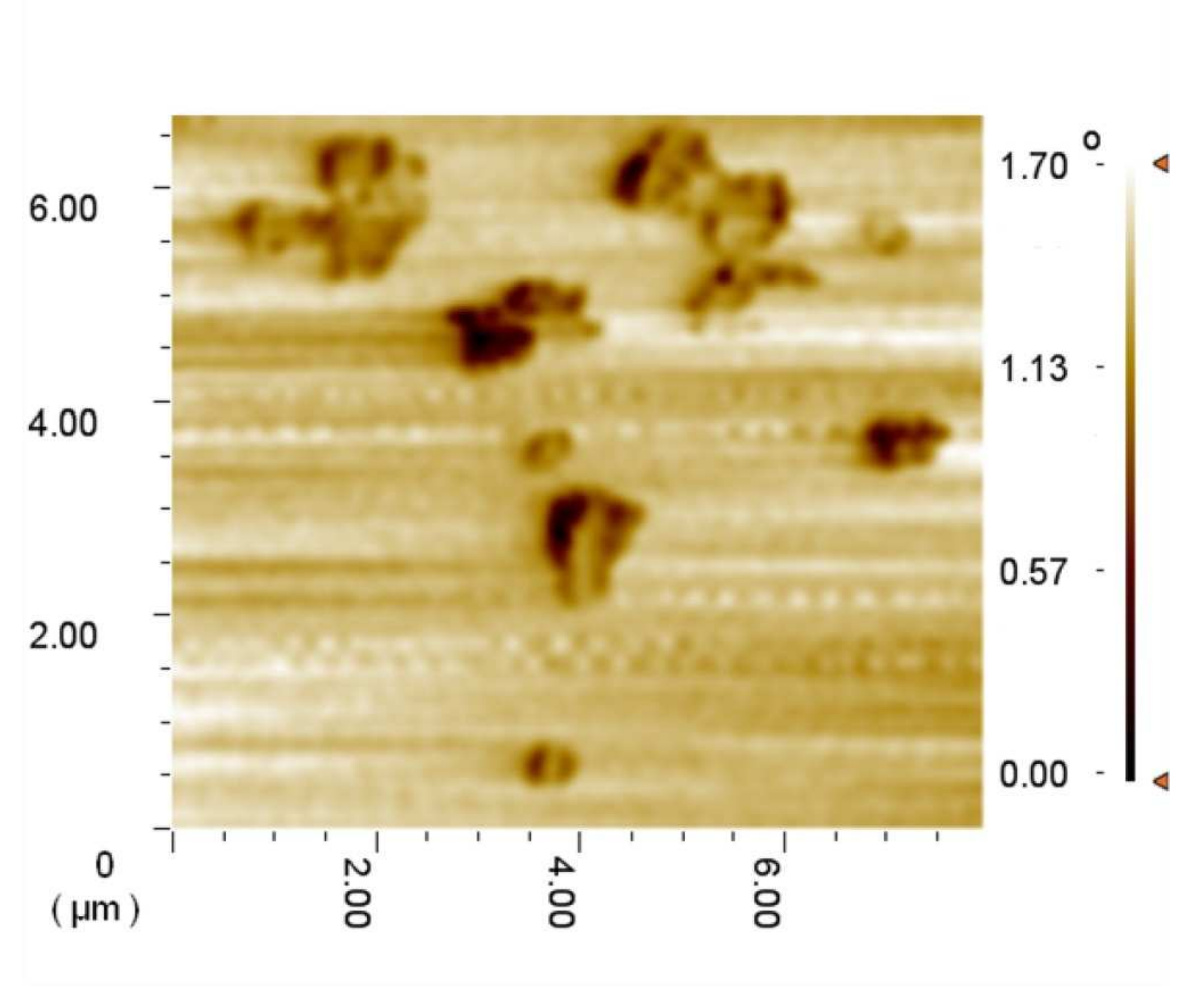}}
   \subfigure[]{\includegraphics[scale=0.25]{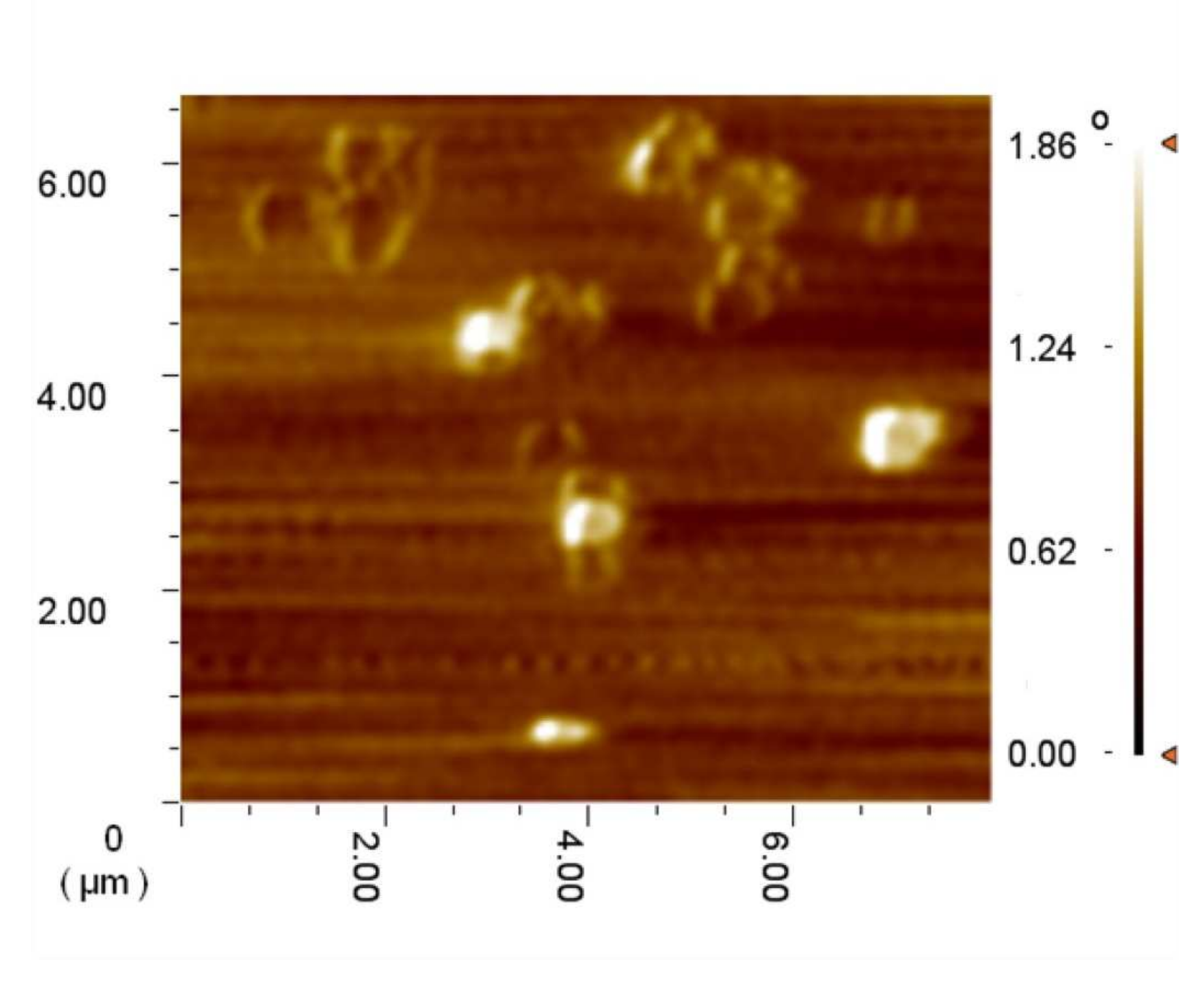}} 
   \end{center}
\caption{The magnetic force microscopy (MFM) images on nanparticle aggregates deposited on glass substrate following dispersion within different liquid media; MFM phase contrast images of several clusters of aggregates over larger area under (a) +20 and (b) -20 kOe field.}
\end{figure}

\begin{figure*}[htp!]
\begin{center}
   \subfigure[]{\includegraphics[scale=0.25]{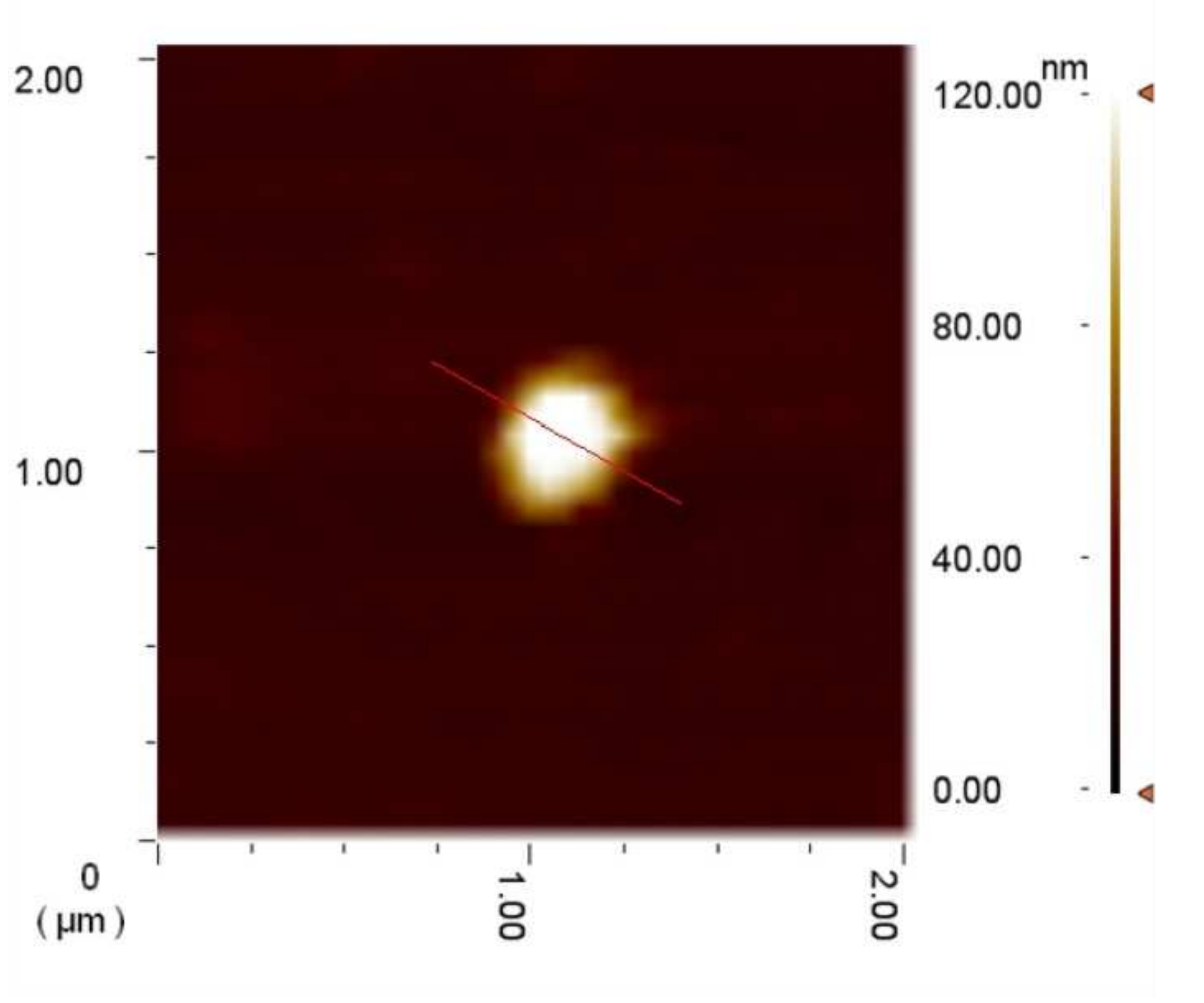}}
   \subfigure[]{\includegraphics[scale=0.25]{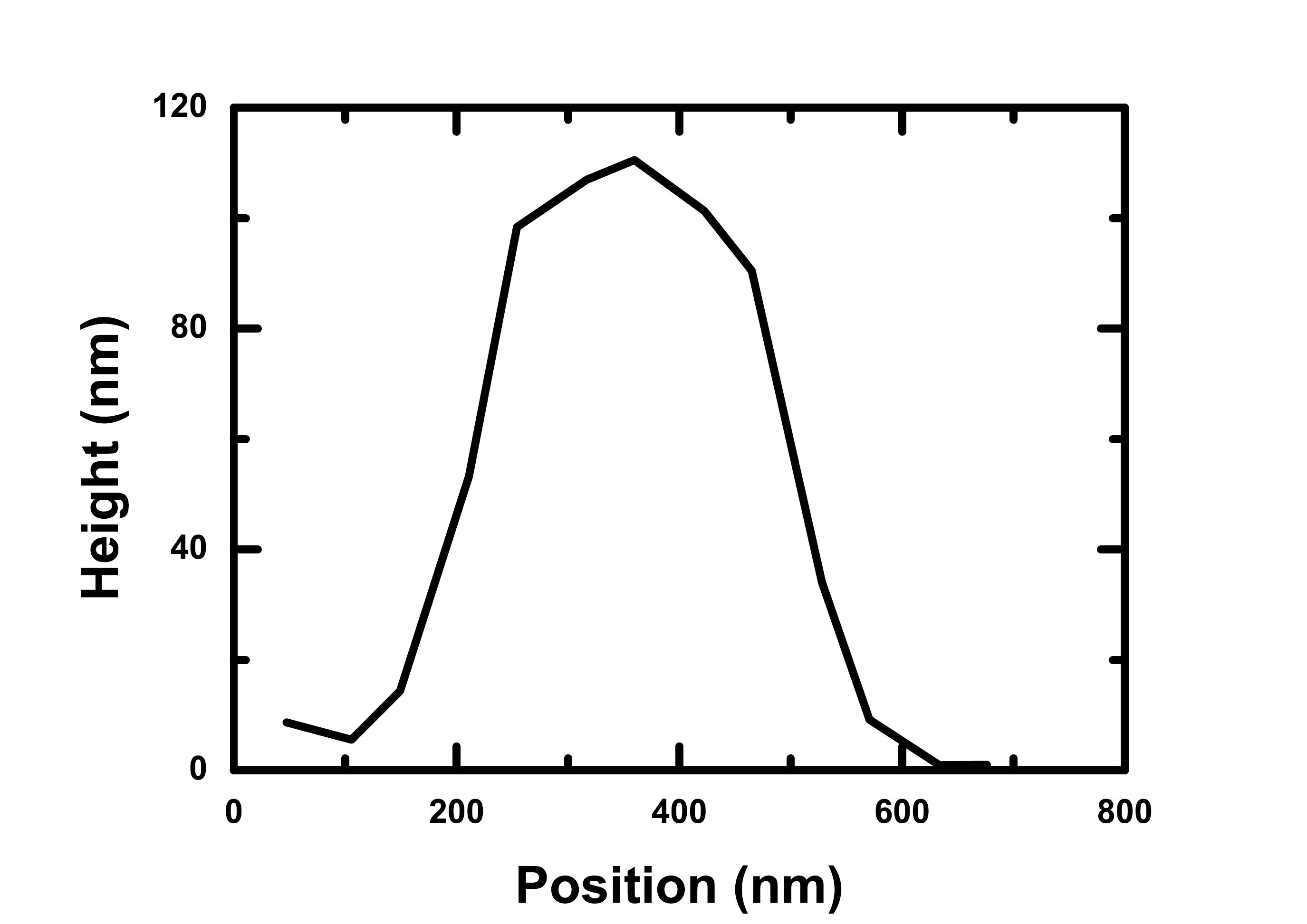}}
   \subfigure[]{\includegraphics[scale=0.25]{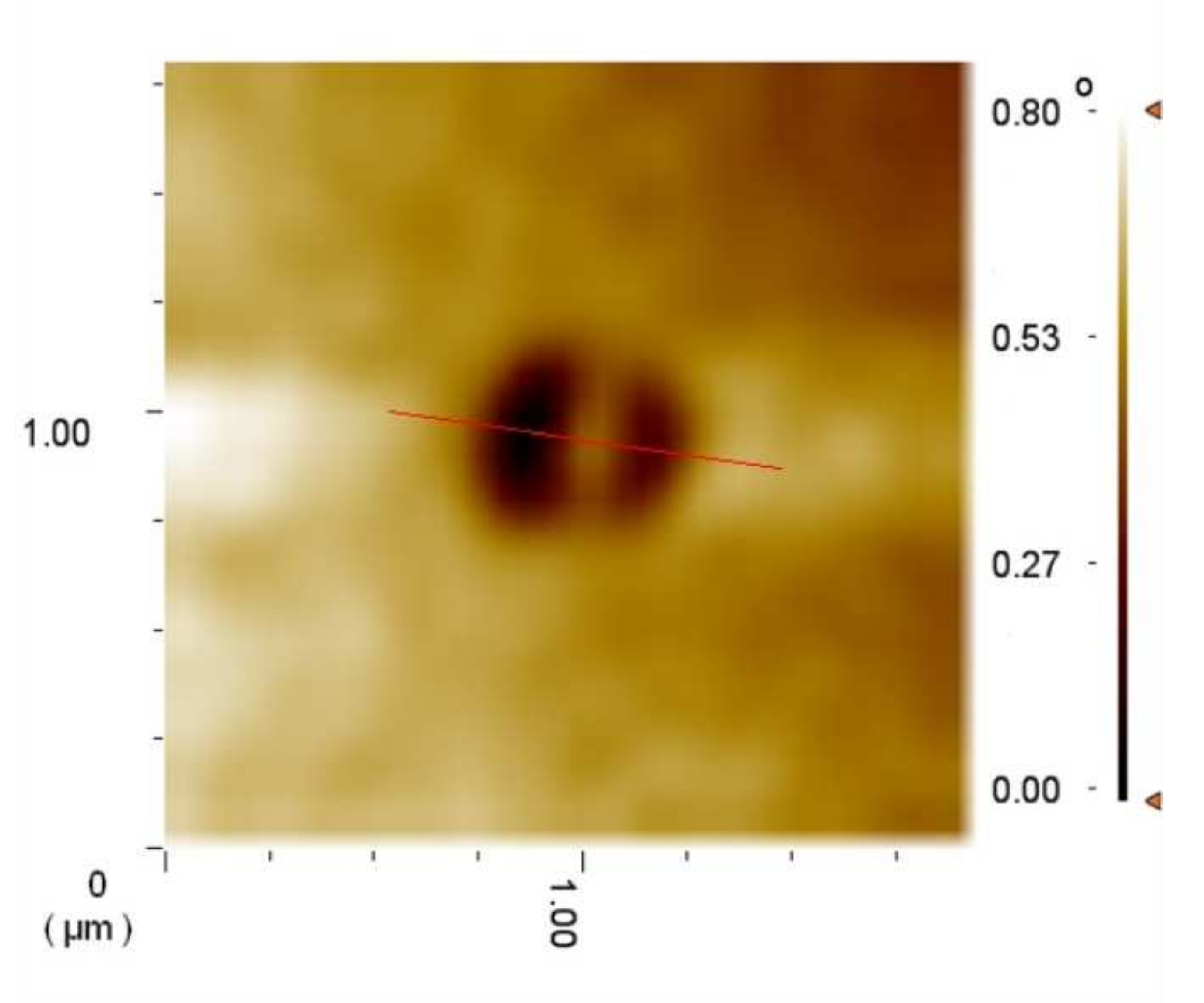}}
   \subfigure[]{\includegraphics[scale=0.25]{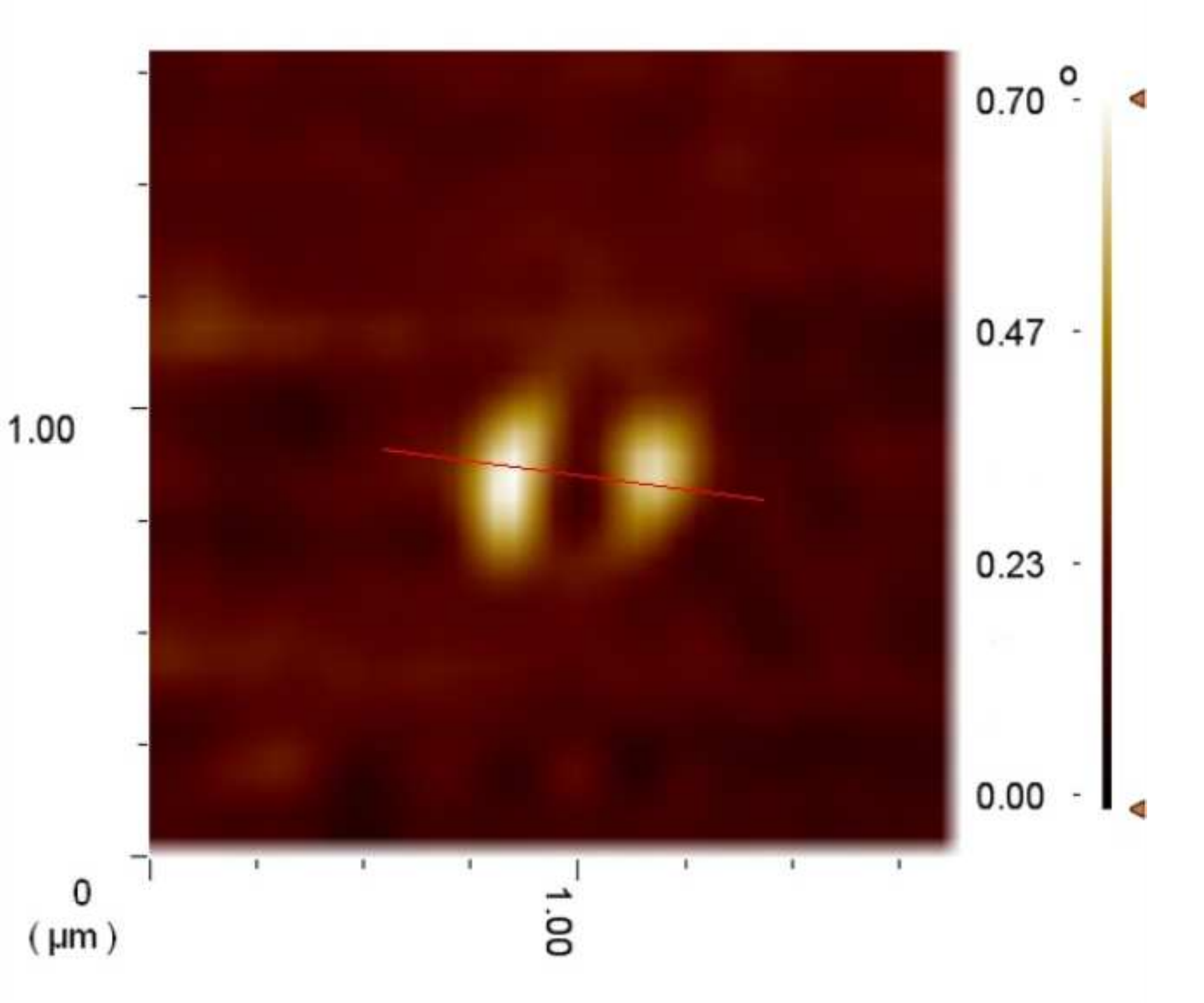}}
   \subfigure[]{\includegraphics[scale=0.25]{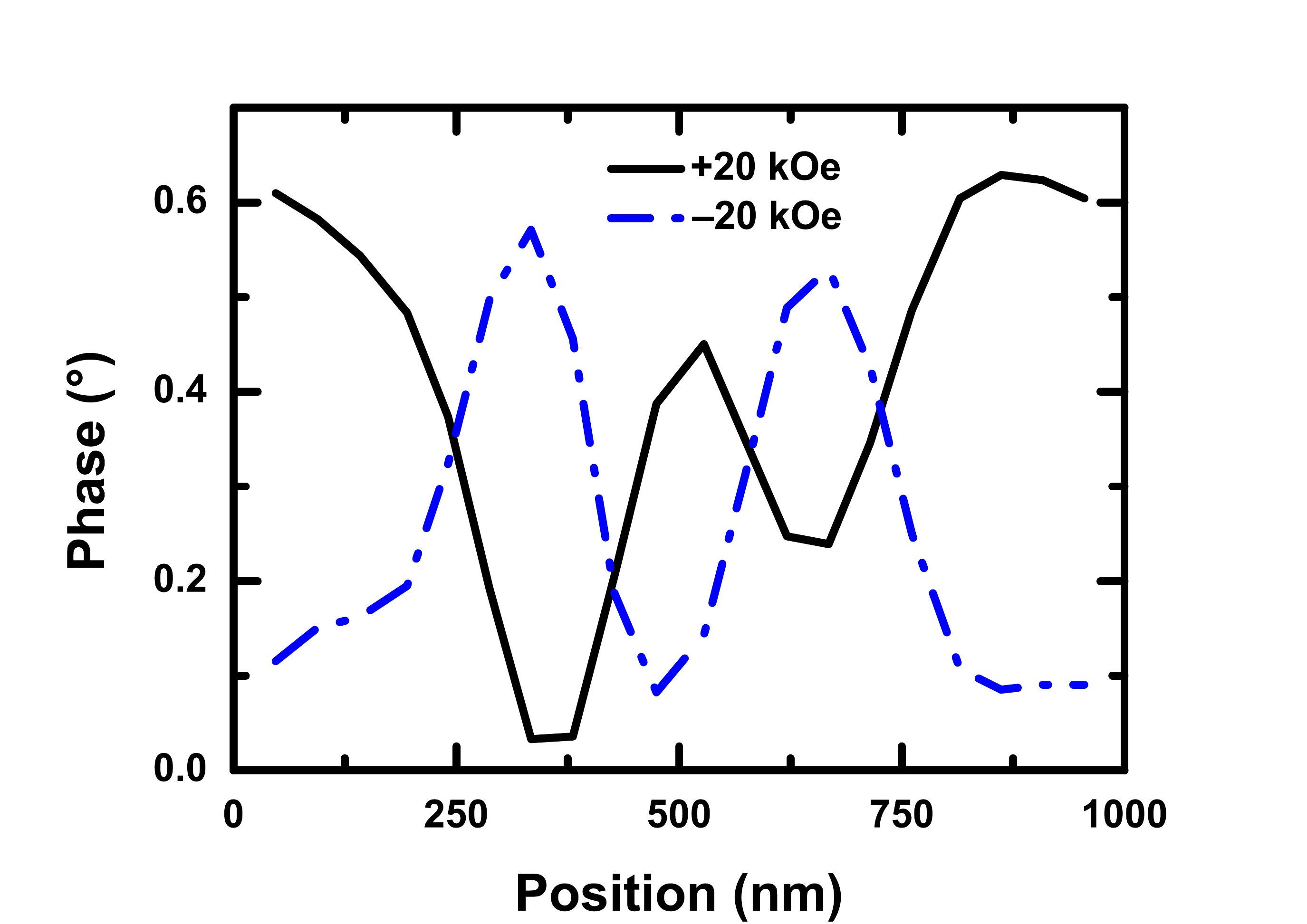}}
   \end{center}
\caption{The magnetic force microscopy (MFM) images for Sample 2; (a), (b) show, respectively, the topography and corresponding crossectional profile; (c),(d),(e) show, respectively, the MFM phase contrast images of Sample 2 under +20 and -20 kOe field and corresponding line profile data. }
\end{figure*}

\begin{figure*}[htp!]
\begin{center}
	\subfigure[]{\includegraphics[scale=0.25]{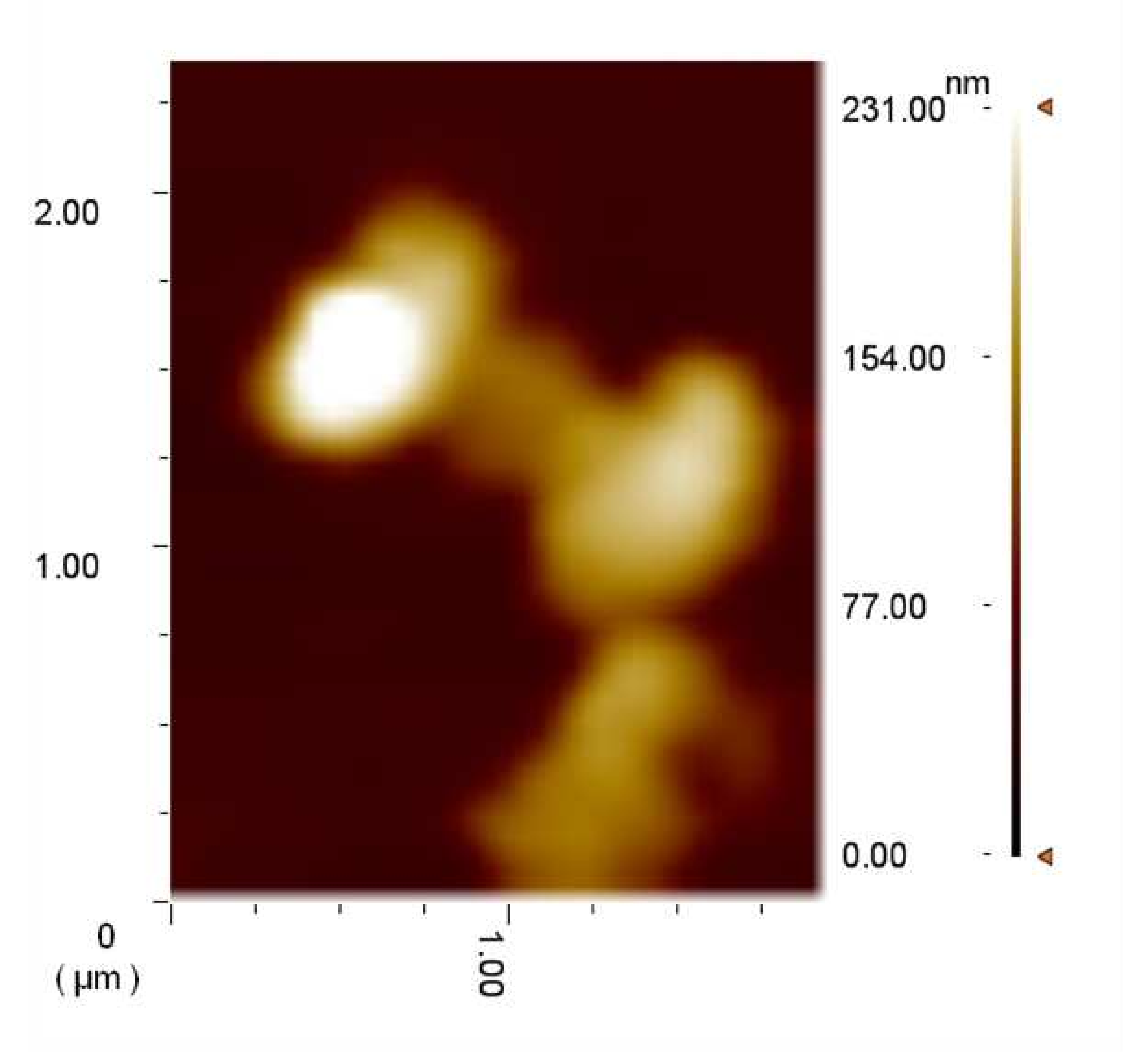}}
	\subfigure[]{\includegraphics[scale=0.25]{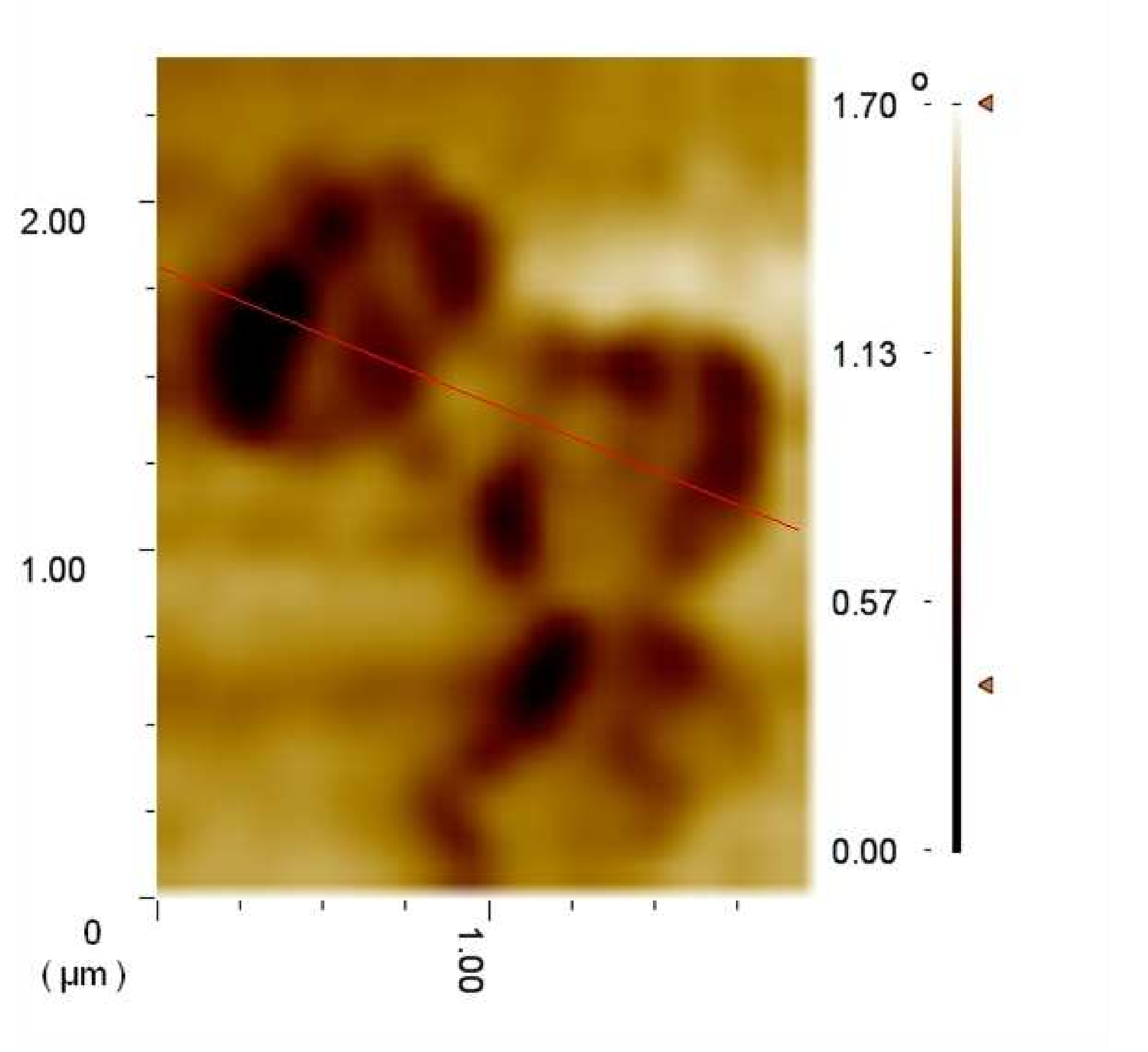}}
	\subfigure[]{\includegraphics[scale=0.25]{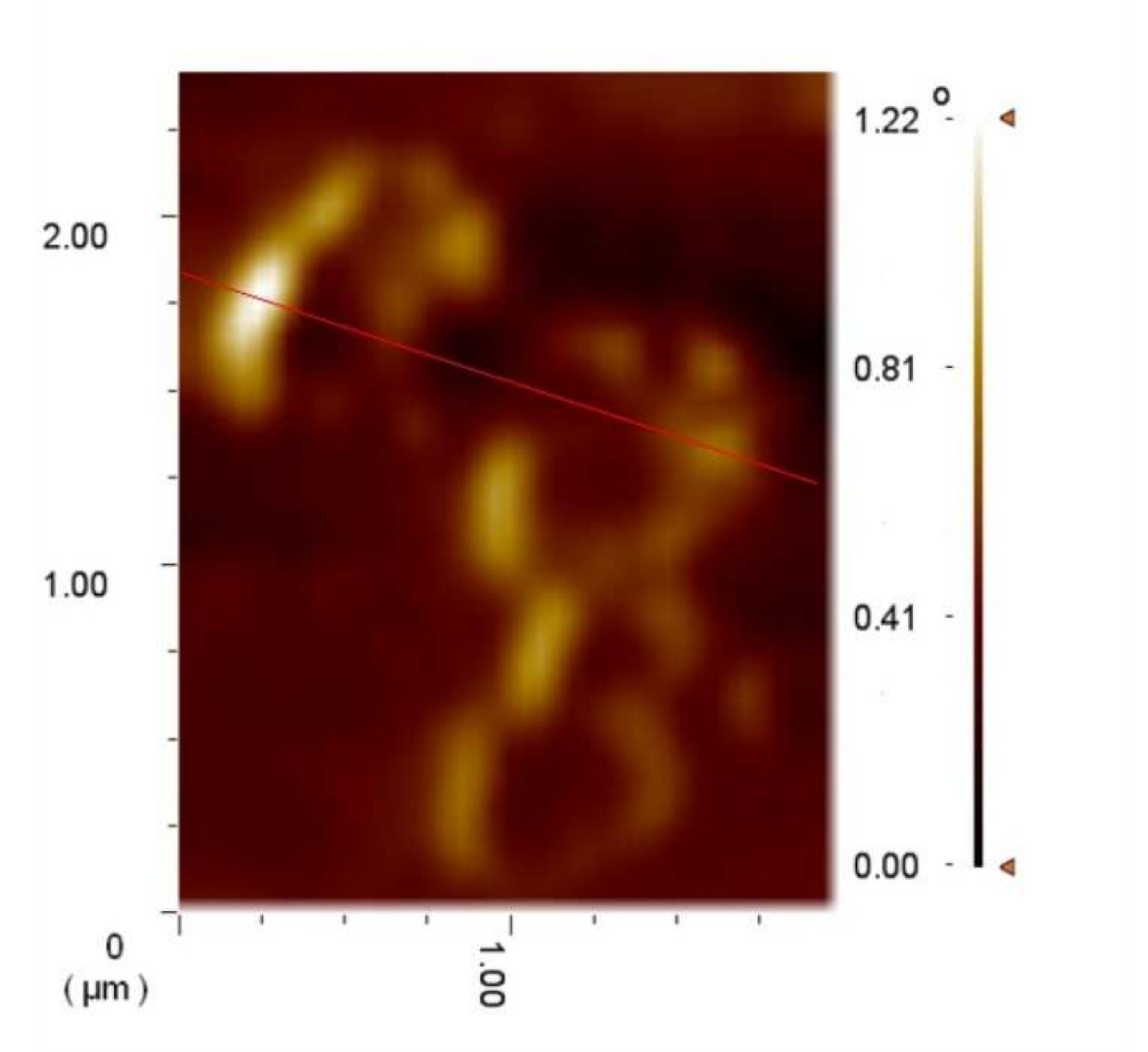}}
	\subfigure[]{\includegraphics[scale=0.25]{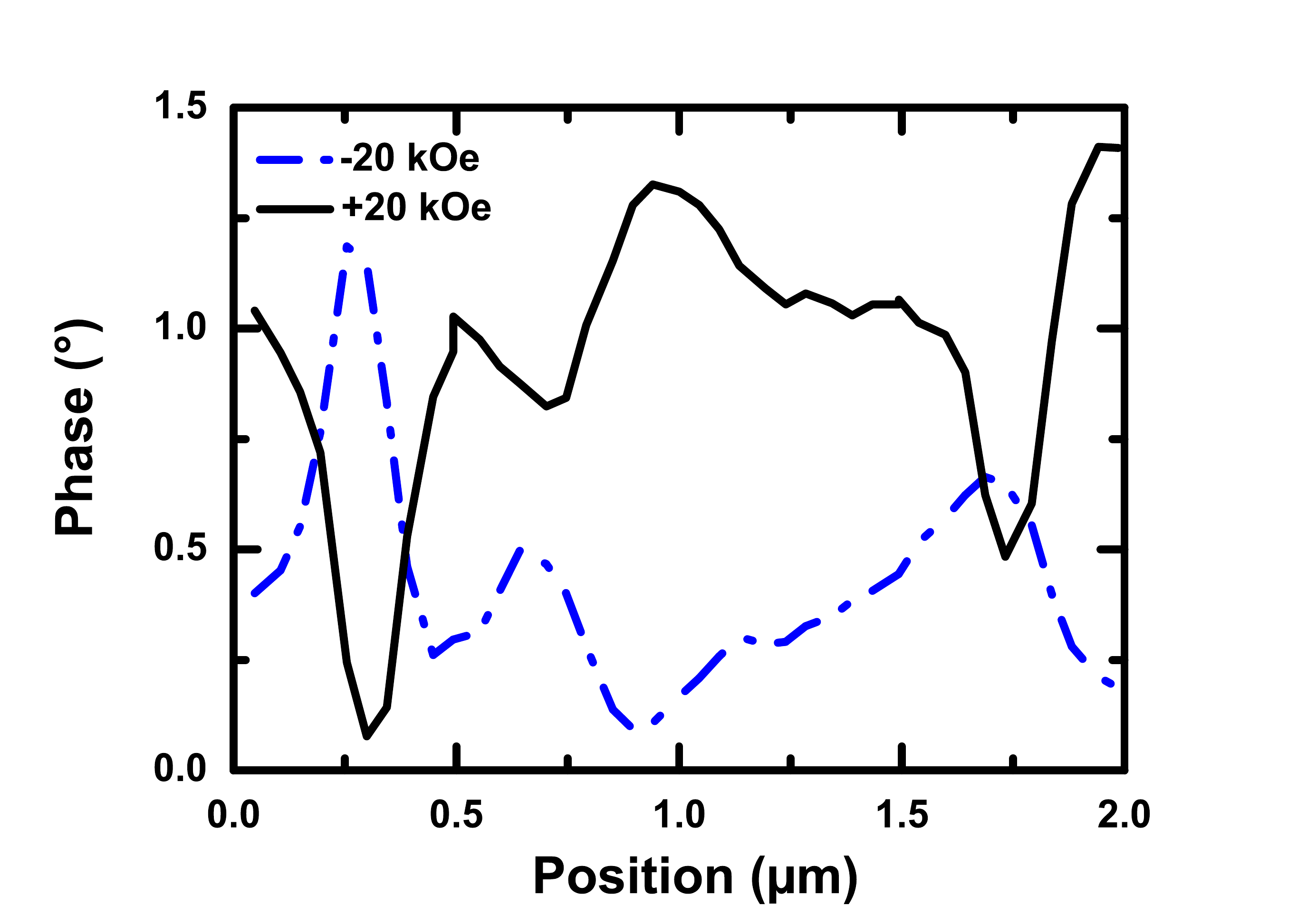}}
   \end{center}
\caption{The magnetic force microscopy (MFM) images for Sample 3; (h) shows the topography; (i),(j),(k) show, respectively, the MFM phase contrast images under +20 and -20 kOe field and corresponding line profile pattern. The domains and their switching under reversal of applied field could be clearly seen in the phase contrast images and corresponding line profile plots. }
\end{figure*}

The magnetic moment ($M$) versus field ($H$) hysteresis loops of isolated nanosized particles as well as self-assembled structures have been measured at room temperature by a Vibrating Sample Magnetometer. We ensured that the samples are not magnetized - even during the synthesis process - prior to any measurement. This is necessary to measure the spontaneous exchange bias. Figure 5 shows the $M-H$ loops measured across $\pm$20 kOe. The total time scale of tracing a complete hysteresis loop was of the order of $\sim$10$^3$-10$^4$s. We also measured the time-scale dependence of the hysteresis loop (see supplementary material). Interestingly, no time dependence could be observed. Therefore, the magnetic order appears to be stable. The magnetic moment versus temperature measurement on the isolated nanoparticles across 300-800 K shows that the magnetic transition $T_N$ is $\sim$ 620 K (see supplementary material). This is consistent with the observation of drop in $T_N$ with the decrease in particle size \cite{Bhattacharya-1}. However, the particles are not superparamagnetic. The room temperature powder neutron diffraction data \cite{Bhattacharya-2} offer evidence of antiferromagnetic order with enhanced canting angle $\sim$6$^o$ (in bulk sample, the canting angle was found to be $\sim$1$^o$ [Ref. 33]). In an earlier work, it has been shown \cite{Sundaresan} that many oxides in nano size exhibit surface ferromagnetism because of surface defects. For BiFeO$_3$ and Bi$_2$Fe$_4$O$_9$ too, surface ferromagnetism with core antiferromagnetic order could be observed in nanosized particles \cite{Tian,Zhang}. In the present case, observation of antiferromagnetic order in powder neutron diffraction data and surface ferromagnetism in magnetic force microscopy (discussed later) point out that the nanoparticles of BiFeO$_3$ possess inverse core-shell spin structure with antiferromagnetic core and ferromagnetic surface. From the measured $M-H$ hysteresis loops, we obtain the coercivity ($H_C$) and exchange bias field ($H_E$) using the relations $H_C$ = ($H_{c1}$ - $H_{c2}$)/2 and $H_E$ = ($H_{c1}$ + $H_{c2}$)/2, where $H_{c1}$ and $H_{c2}$ are the fields corresponding to zero magnetic moment in the forward and reverse branches of the loop \cite{Schuller}. The error bar for the data varies within 1\%-5\%. Following features could be noticed: (i) the coercivity of isolated nanosized particles (Sample 1) is $\sim$250 Oe which is comparable to what has been observed by others \cite{Park} in nanosized BiFeO$_3$; no exchange bias field could be observed; (ii) the nanoparticle aggregate with limited crosslinking (Sample 2), on the other hand, exhibits large coercivity of $\sim$1040 Oe and positive exchange bias field (+$H_E$) of $\sim$875 Oe; (iii) finally, the aggregate with massive crosslinking (Sample 3) is found to exhibit comparable coercivity of $\sim$1100 Oe yet negative exchange bias field (-$H_E$) of $\sim$350 Oe. The loops exhibit negligible vertical shift (compared to what is observed in the case of a minor loop \cite{Nogues-1}) which signifies that they are major and saturation of the ferromagnetic component could be achieved within the field regime applied ($\pm$20 kOe). Of course, complete saturation of the overall magnetization could not be observed as the volume fraction of the ferromagnetic component is relatively small. Similar observations have been made by others as well \cite{Wang,Kleemann}. Importantly, the loops also exhibit reversibility of magnetization at the high field regime which signifies that they are major \cite{Goswami-3}. The volume fraction of the ferromagnetic component appears to be higher in the case of Sample 3. The loop shape in this case is possibly influenced by combined effect of the following factors: (i) enhancement of ferromagnetism and that of domain size (discussed later), (ii) and decrease in hysteretic spin fraction as a result of both finite surface spin pinning (nonhysteretic part) and increase in domain size which leads to decrease in surface area to volume ratio. Therefore, the coercivity is smaller than what is expected for a sample with enhanced ferromagnetic volume fraction. Of course, in comparison to the coercivity observed in the case of the aggregate with limited crosslinking (Sample 2), the coercivity is slightly higher in Sample 3. Observations such as (i) enhancement of coercivity by more than a factor of 4 and (ii) spontaneously developed \cite{Saha,Wang} large exchange bias field ($H_E$) in nanoparticle aggregates as well as (iii) structure-dependent switching of $H_E$ from negative to positive with larger $\mid +H_E \mid$ are indeed remarkable. As against this spontaneous switching of exchange bias, earlier works \cite{Goswami-4,Goswami-5} on nanocomposites of BiFeO$_3$/Bi$_2$Fe$_4$O$_9$ showed such switching to be resulting from switch in the direction of the first applied field - from positive to negative - while tracing the complete hysteresis loop. It is important to mention here that we have ensured that at the application of first field for measurement of the hysteresis loop (measurement of initial magnetization branch) the magnetic moment in both the cases of self-assembled patterns was nearly identical. This signifies nearly identical initial states in both the cases. The isolated nanoparticles with core antiferromagnetism and surface ferromagnetism, of course, do not exhibit detectable exchange bias. The magnitude of $H_E$ in `nanoparticle aggregates' studied here, though large, is, of course, not quite as high as $\sim$10-30 kOe observed in natural mineral (FeTiO$_3$-bearing Fe$_2$O$_3$) \cite{McEnroe} and a family of Heusler alloy systems \cite{Nayak}.

\begin{figure}[htp!]
\begin{center}
   \subfigure[]{\includegraphics[scale=0.25]{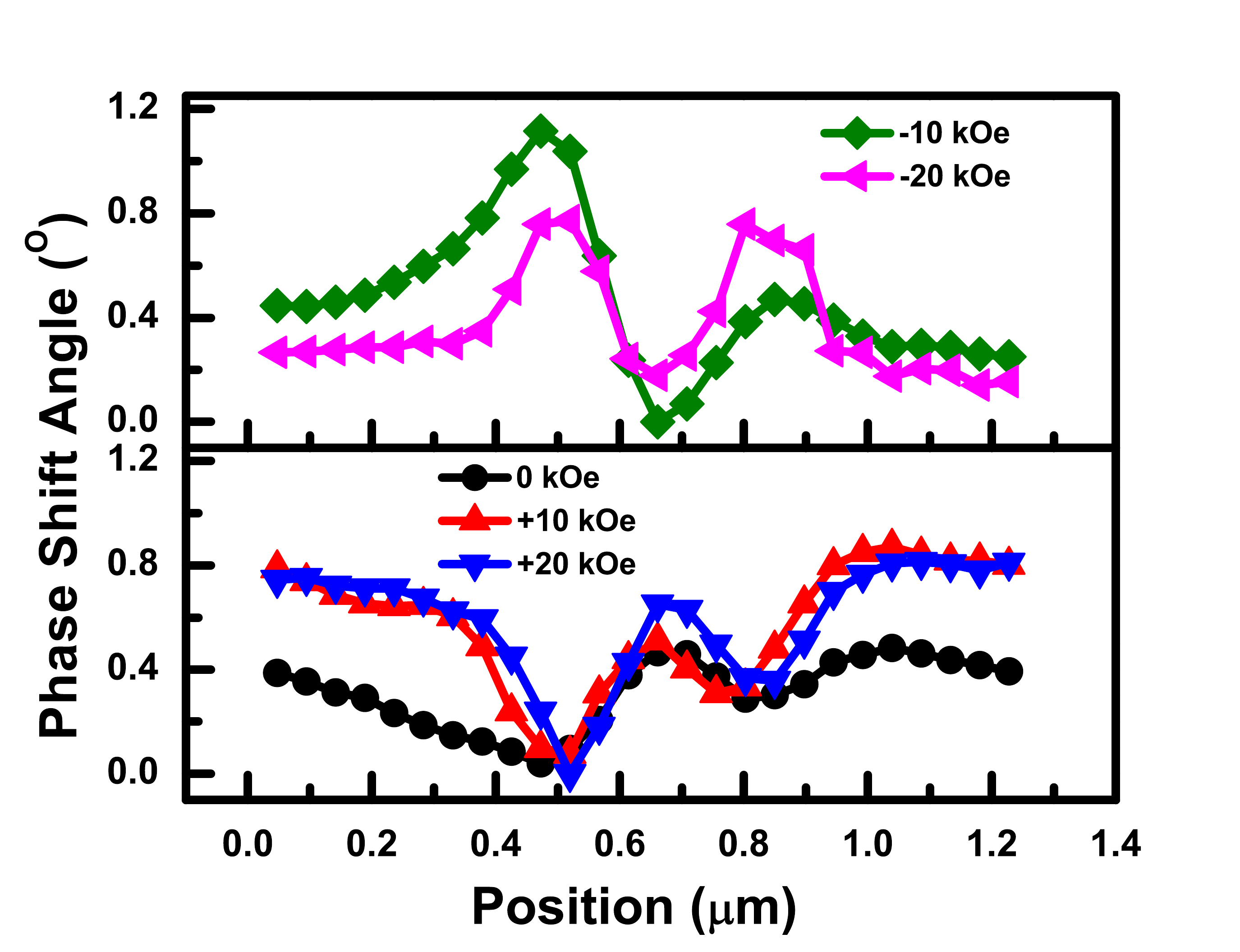}}
   \subfigure[]{\includegraphics[scale=0.25]{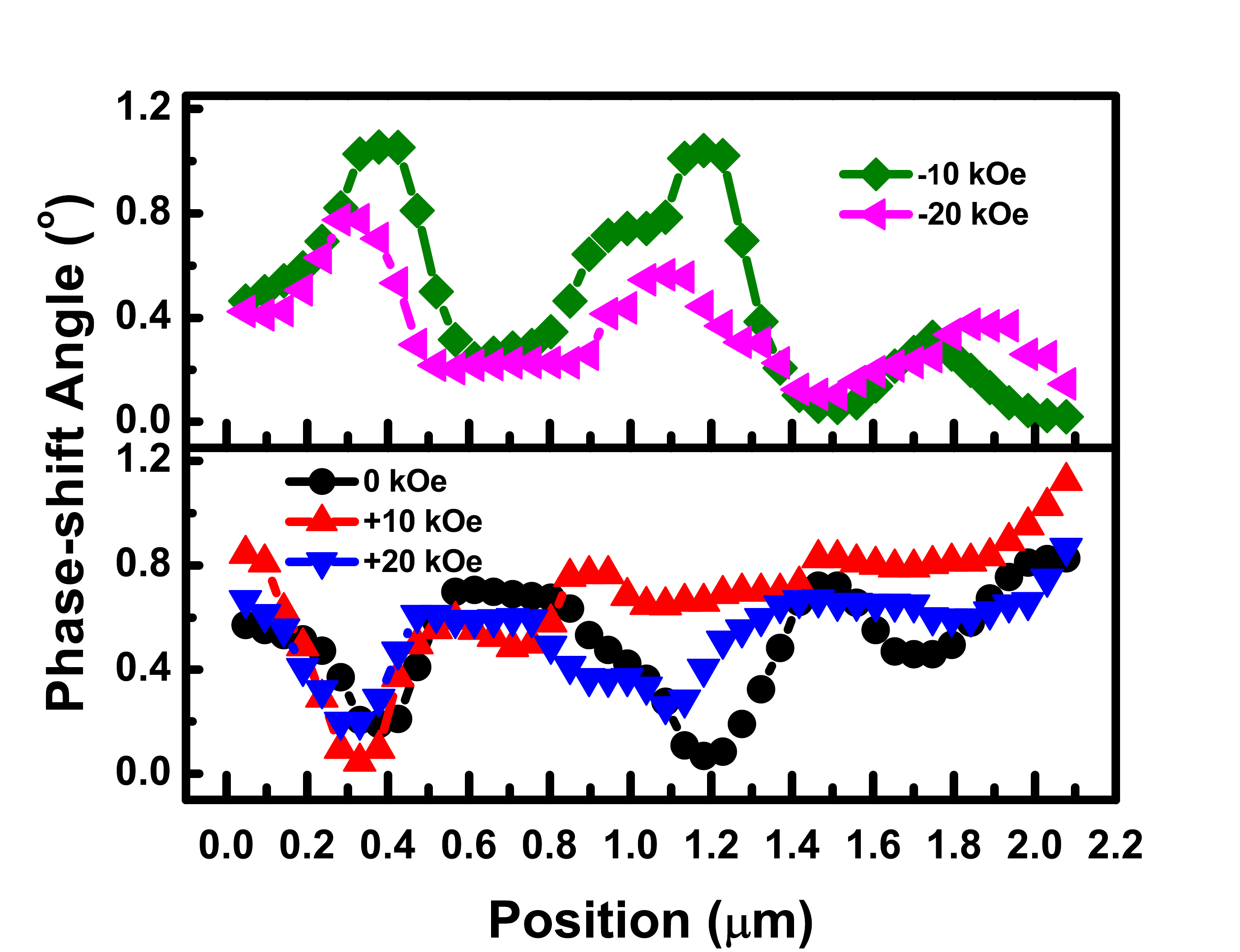}} 
   \end{center}
\caption{The phase-shift angle versus position data at different fields for (a) Sample 2 and (b) Sample 3. }
\end{figure}

\begin{figure*}[htp!]
\begin{center}
   \subfigure[]{\includegraphics[scale=0.15]{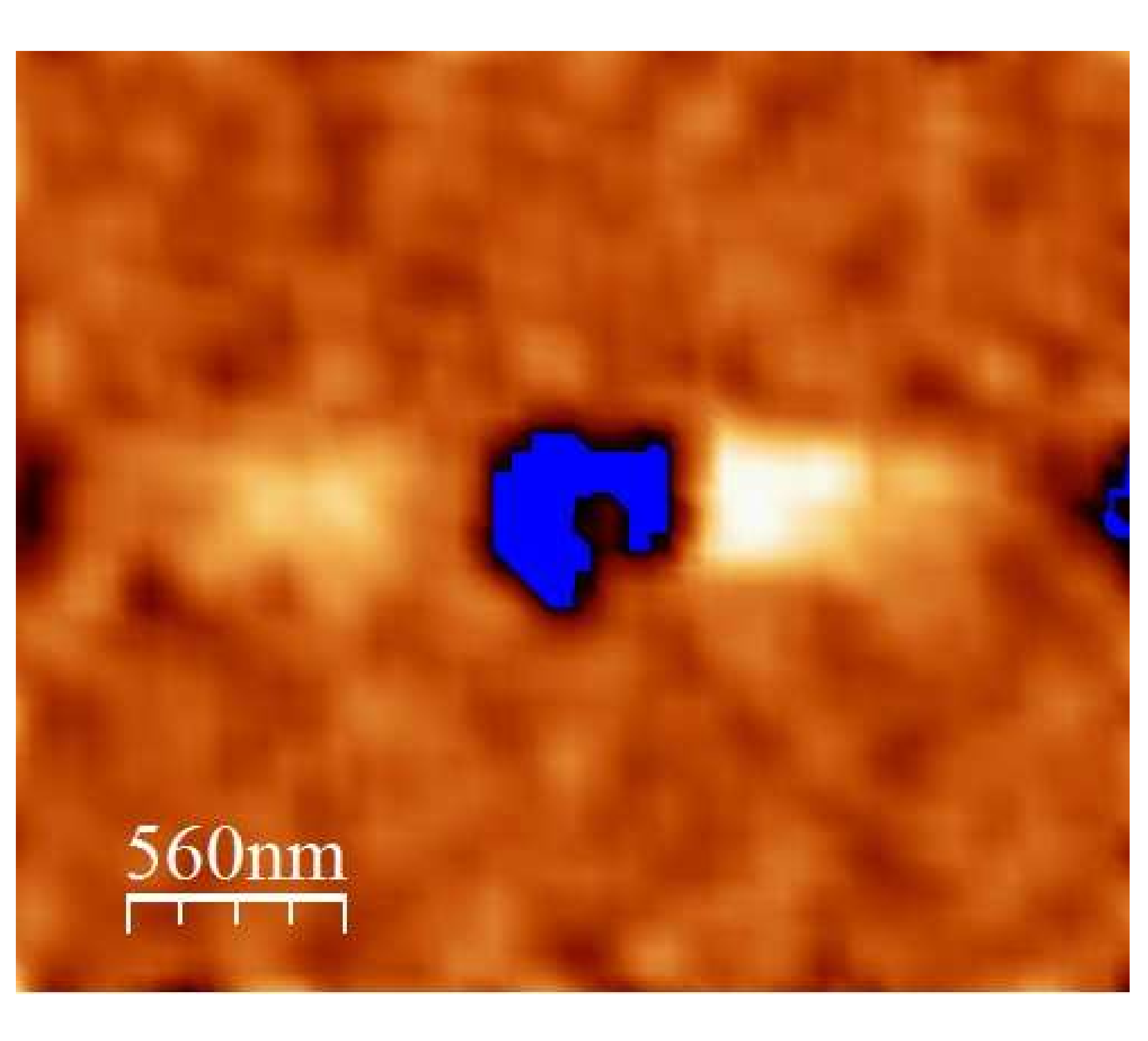}}
   \subfigure[]{\includegraphics[scale=0.15]{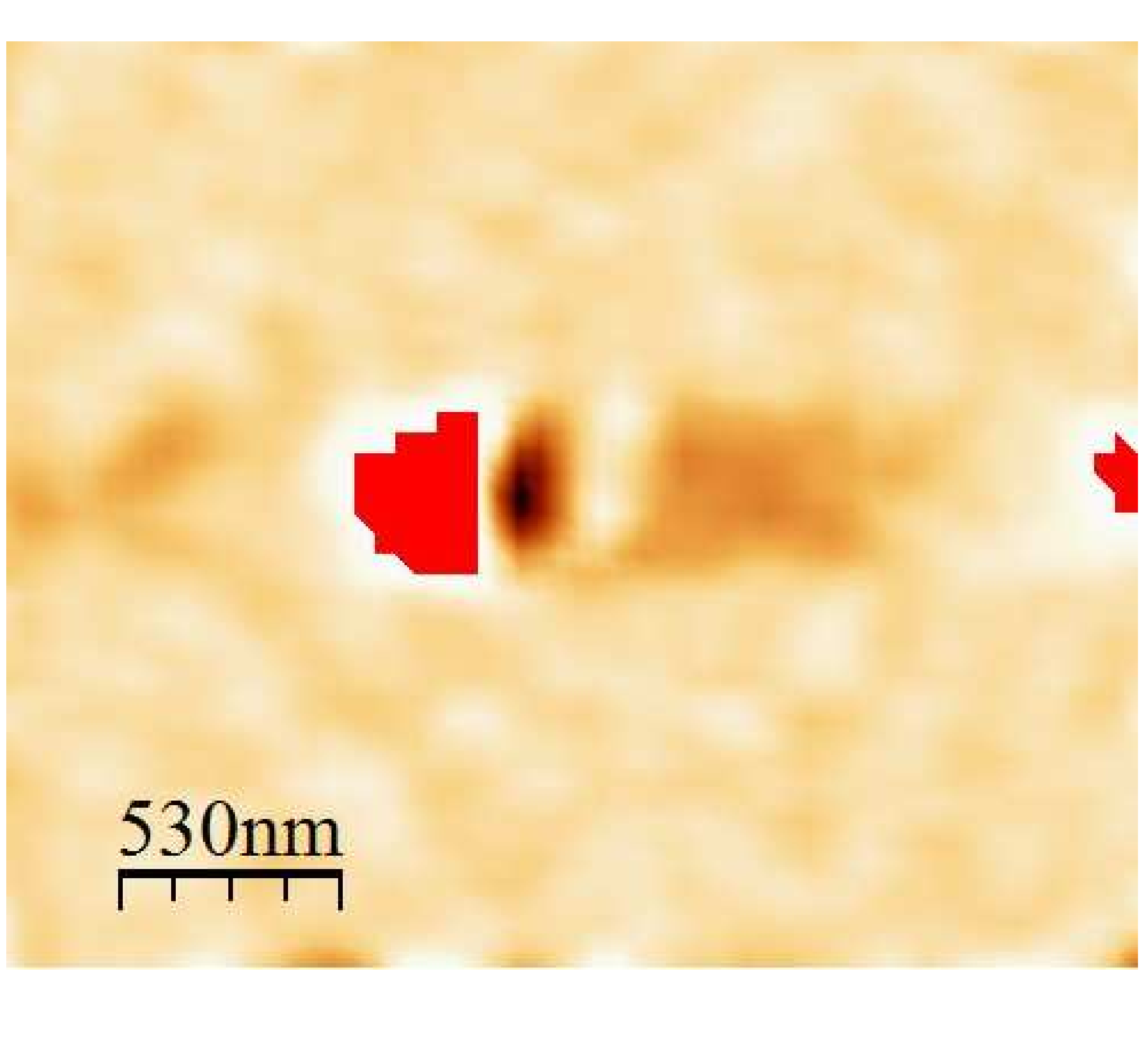}}
   \subfigure[]{\includegraphics[scale=0.15]{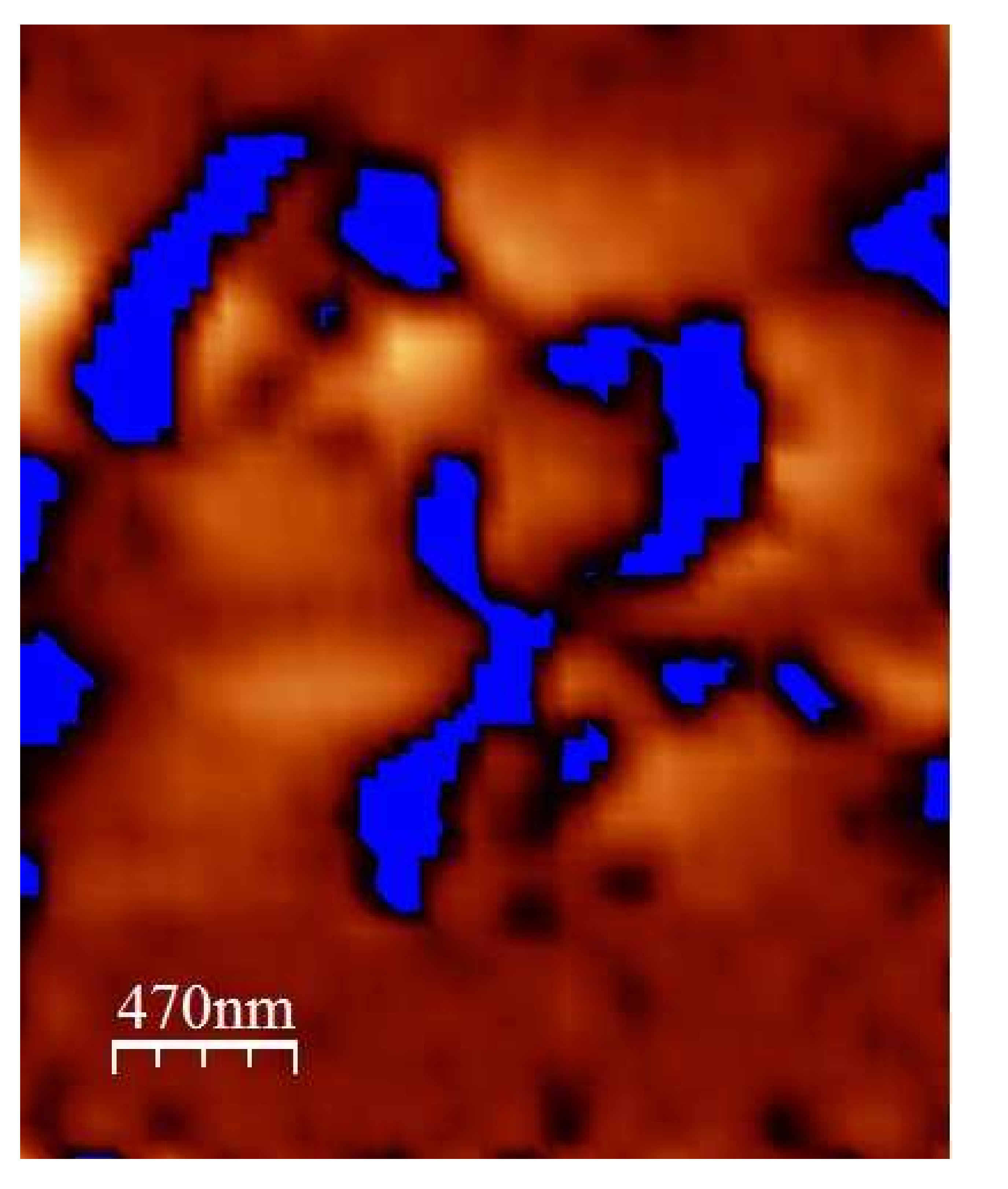}}
   \subfigure[]{\includegraphics[scale=0.15]{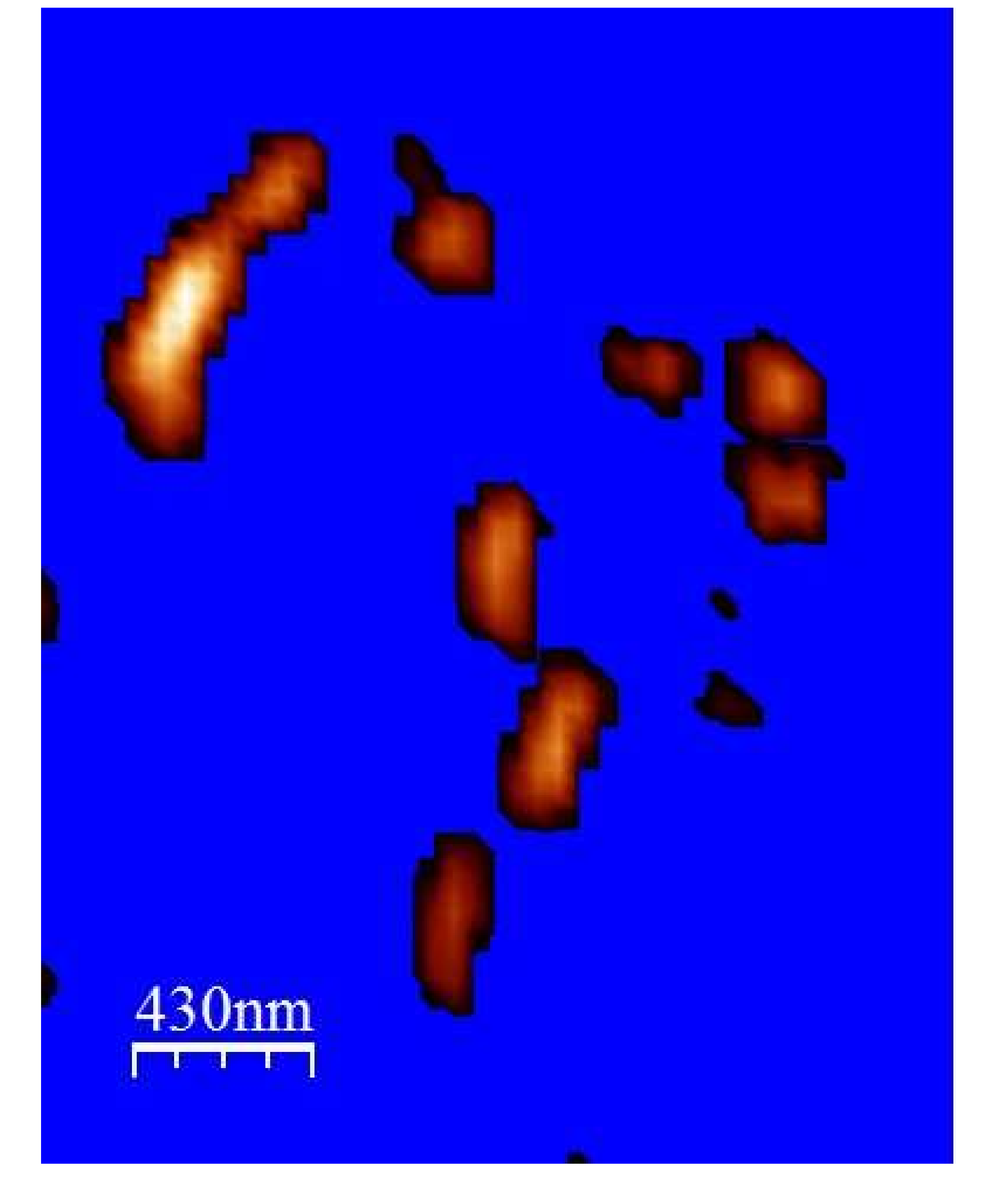}}
   \end{center}
\caption{The processed MFM phase contrast images (a), (b) show, respectively, the domain volume switching in Sample 2 under +20 and -20 kOe while (c), (d) show, respectively, the domain volume switching in Sample 3 under +20 and -20 kOe. }
\end{figure*}

\begin{figure}[htp!]
\begin{center}
   \subfigure[]{\includegraphics[scale=0.25]{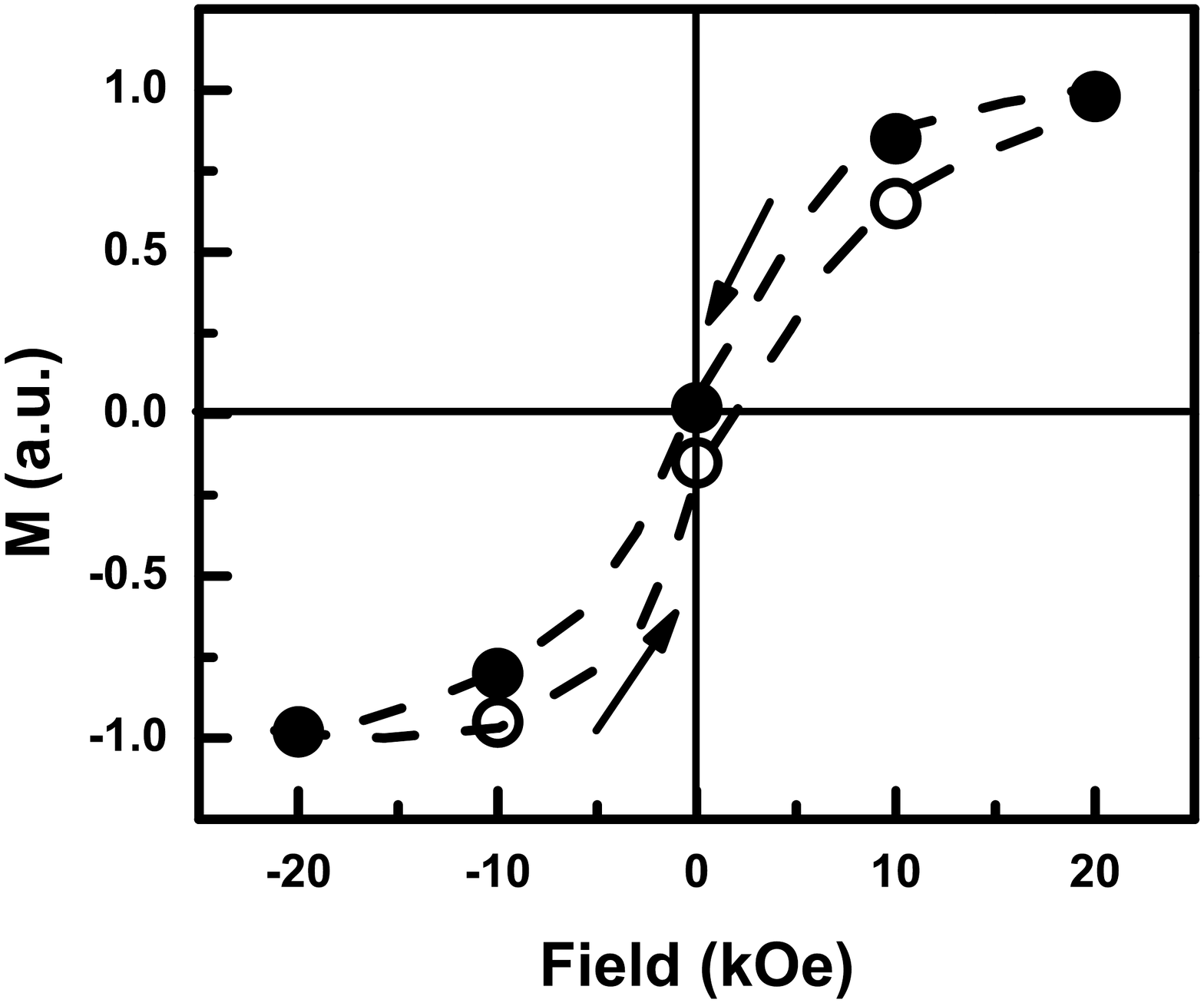}}
   \subfigure[]{\includegraphics[scale=0.25]{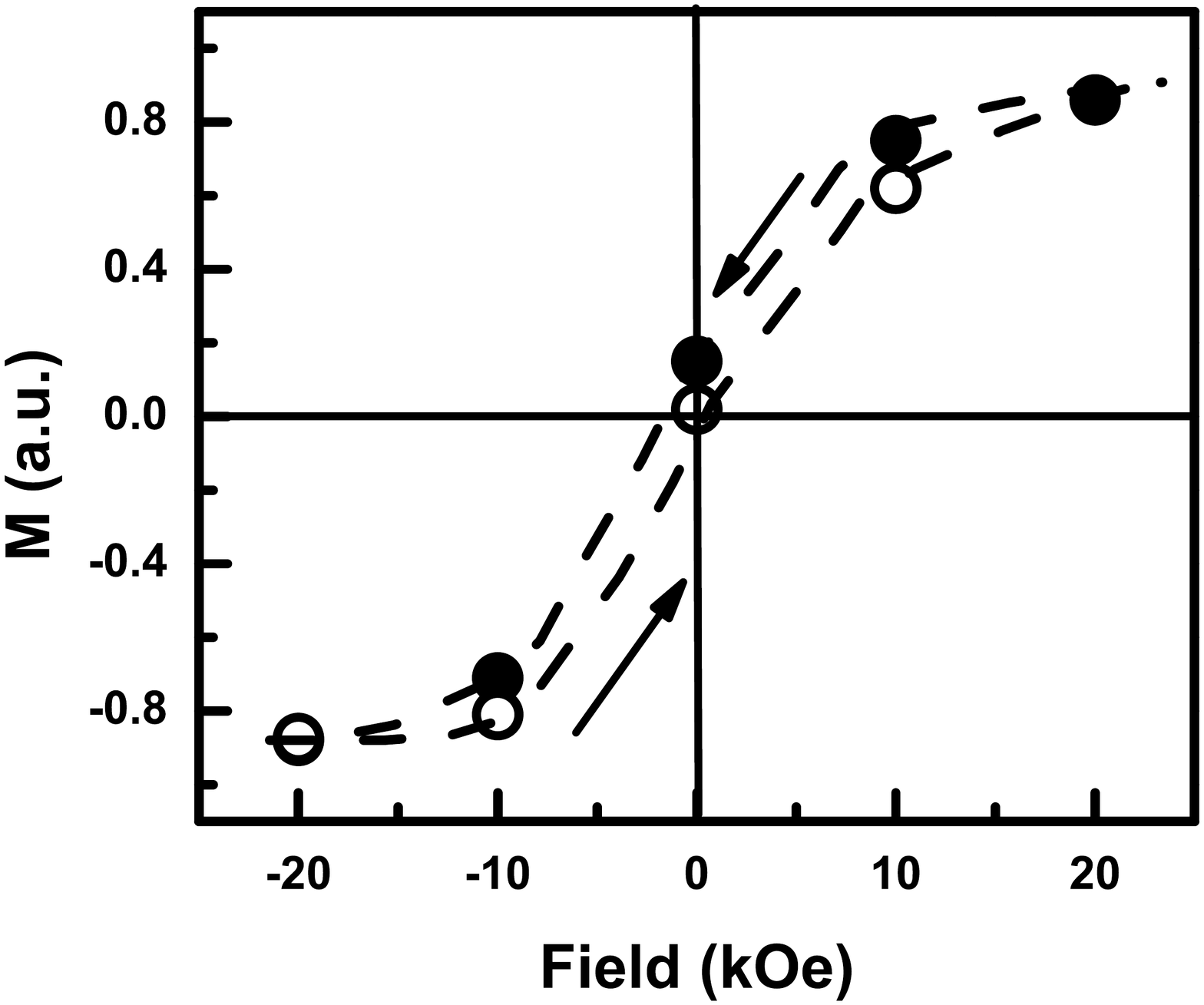}}
   \end{center}
\caption{The magnetic hysteresis loops obtained from the data of 'switched domain volume' under field sweep for (a) Sample 2 and (b) Sample 3.}
\end{figure}

In Figure 6, we show the typical MFM phase contrast images captured across several nanoparticle aggregates over a large area. The Figures 6(a) and 6(b) show, respectively, the images under +20 kOe and -20 kOe. The Figure 7 shows the MFM images for Sample 2. While Fig. 7(a) shows the topography of Sample 2 [with corresponding cross-sectional profile data in Fig. 7(b)], the MFM phase contrast images for Sample 2 under +20 and -20 kOe field are shown, respectively, in Figs. 7(c) and 7(d). The Figure 7(e), finally, shows the corresponding line-profile data for the images of Figs. 7(b) and 7(c). Likewise, Figure 8(a) shows the topography of Sample 3 and Figs. 8(b), (c) show, respectively, the MFM phase contrast images for Sample 3 under +20 and -20 kOe field. The corresponding line-profile data are shown in Fig. 8(d). Since results presented in Figs. 7 and 8 are, specifically, for Samples 2 and 3, respectively, we restrict ourselves in analysing these results alone in what follows. It is important to mention here that even though the cantilever moment should reverse its direction when the direction of the applied field is reversed, we ensured that this does not influence the image. In the case of influence of the reversal of cantilever moment, the images under negative applied field would have been exactly the reverse contrast images of those obtained under positive applied field even if the domain reversal does not occur in the sample under the applied field. This has not been observed in the present case. From the $M-H$ hysteresis loop measurement, it has already been noticed that indeed the magnetic moment of the sample reverses its sign with the change in the sign of applied field. The coercivity of the nanochains is much higher than that of the cantilever. Therefore, we conclude that the MFM captures the actual reversal of domains of the samples (and not of the cantilevers) under reversal of the direction of applied field. The change in background color of the `processed MFM phase contrast images' (obtained under different magnetic fields including reversal of the field) could possibly originate from image processing by SPM software in which `absolute mode' was used to show the amplitude of phase shift. For this mode of scaling, the dark color is coded zero and the bright color is coded maximum amplitude value for the two cases. The colors are assigned from this image construction perspective. The initial phase of the cantilever, which is different for these two cases, may give a different background color. The main focus for an MFM image is, of course, the contrast in the `zone of interest' with its amplitude (i.e, magnitude of phase shift) and not the background color which has no physical meaning as far as information on domain switching is concerned. Following features are noticeable from the MFM images: (i) the line profile analysis shows the domain size to be varying within 100-150 nm and reasonably uniform for Sample 2; (ii) however, for Sample 3, the domain size appears to have variation over a scale of 100-250 nm; (iii) consistent with the observations made by others within such array of nanoparticles \cite{Alivisatos}, domain size turns out to be greater than the particle size which signfies formation of a domain across several particles (in fact, domain size turns out to be comparable with the total size of 7-10 nanoparticles); (iv) in both the cases, of course, complete domain switching could be observed under sweeping magnetic field from +20 kOe to -20 kOe. It is possible to infer that because of the presence of bigger domains, Sample 3 exhibits smaller unidirectional anisotropy and hence $H_E$. More MFM images are available in the supplementary material. 

We have used the phase-shift angle profile data to determine the entire pattern of domain switching characteristics under sweeping magnetic field for both the samples (Sample 2 and 3). The field is swept as per the following sequence: +20 kOe $\rightarrow$ +10 kOe $\rightarrow$ 0 $\rightarrow$ -10 kOe $\rightarrow$ -20 kOe $\rightarrow$ -10 kOe $\rightarrow$ 0 $\rightarrow$ +10 kOe $\rightarrow$ +20 kOe. More MFM images for the samples at different fields are shown in the supplementary material. In Figures 9(a) and 9(b), respectively, we show the phase-shift angle versus position data coresponding to the line profiles taken on the MFM phase contrast images for Sample 2 and 3. The change in the patterns under different applied magnetic fields including the reversal of the field is noticeable. The Figures 10(a), 10(b), 10(c), and 10(d), on the other hand, show the processed `MFM phase contrast images' where the fraction of the `switched domains' under +20 kOe and -20 kOe field are shown in different colors; blue color signifies domain volume switched under +20 kOe while the red color signifies the domain volume switched under -20 kOe. The calculation of the `switched domain volume' has been carried out by processing the MFM images using the image processing software WSxM. The background of the image has been identified appropriately. Using the phase-shift scale of the phase contrast image, the volume of the sample exhibiting switch or change in phase contrast with respect to the background was determined. The ratio of the switched volume with respect to the total volume, then, provides the fraction of the `switched volume of domains'. The change in the fraction is mapped as a function of applied field. This estimation complements the estimation of `switched volume of domains' obtained from area under the line profile plot. The line profile across a particular line on the MFM phase contrast image shows the domain pattern under a given applied field. Several line profiles were used across the area under focus to obtain an average estimate of the `switched domain volume'. The `switched domain volume' thus obtained helps in mapping its variation as a function of applied field. The information collected from both the analyses helps in generating the magnetic hysteresis plots shown in Figs. 11(a) and 11(b). The plot of volume of switched domains as a function of applied field replicates, qualitatively, the magnetic hysteresis plot obtained from magnetic measurements \cite{Zhu}. It is remarkable to notice that, as observed in the magnetic hysteresis plots, plot of `switched domain volume' versus `field' also clearly reflects the presence of exchange bias field and its switching from positive to negative depending on the structure of nanoparticle aggregates. Therefore, both the magnetic measurements as well as imaging of magnetic domain switching characteristics offer clear evidence of presence of sizable exchange bias field. More interestingly, the exchange bias field appears to switch from negative to positive depending on the structure of the aggregates. 

All these results could be rationalized by resorting to the concept of surface spin pinning effect in nanoparticles \cite{Sahoo}. Using Monte Carlo simulations, the magnetic hysteresis loops for different nanoparticle aggregate structures have been obtained. The simulations were carried out on two structures - a limited crosslinked structure with nine nanoparticles and a massive crosslinked compact structure of nine nanoparticles. The parameters used in the model are the particle size $(R)$, core size $(R_c)$, width of overlapping area $K$, the three exchange coupling constants; $J_c$, $J_{sh}$, $J_{int}$, pinning density $(\eta)$, fraction of pinned spins oriented preferably along easy-axis. Experimentally, we have seen that the dispersing solvent is playing an important role in forming different patterns. The earlier work \cite{Berkowitz} has already pointed out that organic solvent plays an important role in inducing surface pinning of spins in ferrite materials. In our case, the surface spin pinning occurs because of functionalization of the surface by organic solvents (kerosene, BBP). This, in turn, yields a finite exchange bias or asymmetry in the hysteresis loop.

\begin{figure}[h!]
	\centering
		\includegraphics[scale=0.20]{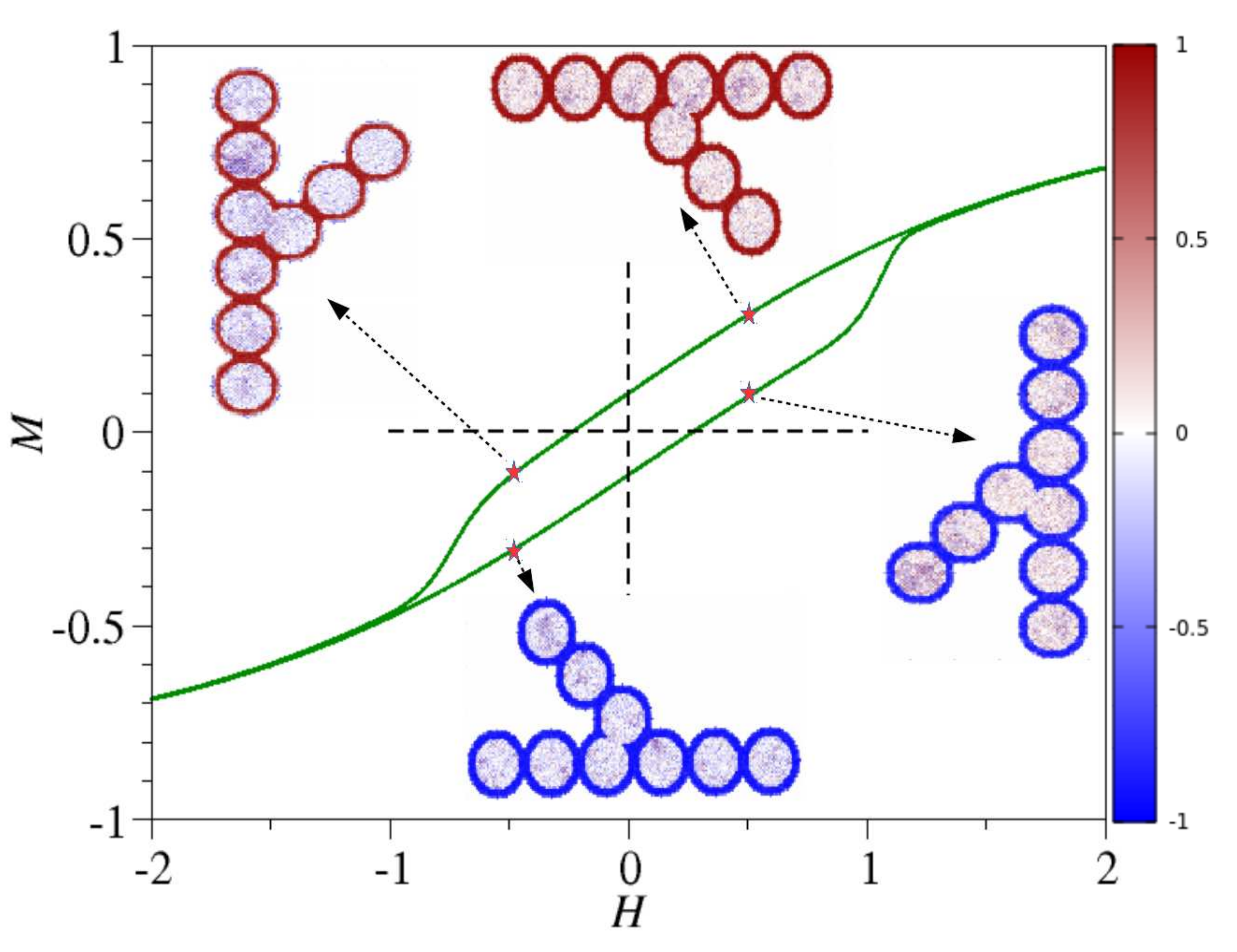}
		\caption{Hysteresis loop for nine crosslinked nanoparticles, each of size $R=16$ having an antiferromagnetic core of radius $R_c=12$; the overlap size of any consecutive particle is $K$=3. A cross-sectional view (at $z=R$)  of the local magnetic  moment (averages over $100$ samples) is shown  at $H=\pm 0.5$ in the forward and backward directions. The interaction parameters are $J_c=-0.5, J_{sh}=1.0, J_{int}=1.0$ and the pinning parameters are $\eta =0.4, r=0.2.$
		 }
		\label{fig:crsA.pdf}
\end{figure}

\begin{figure}[h!]
	\centering
		\includegraphics[scale=0.20]{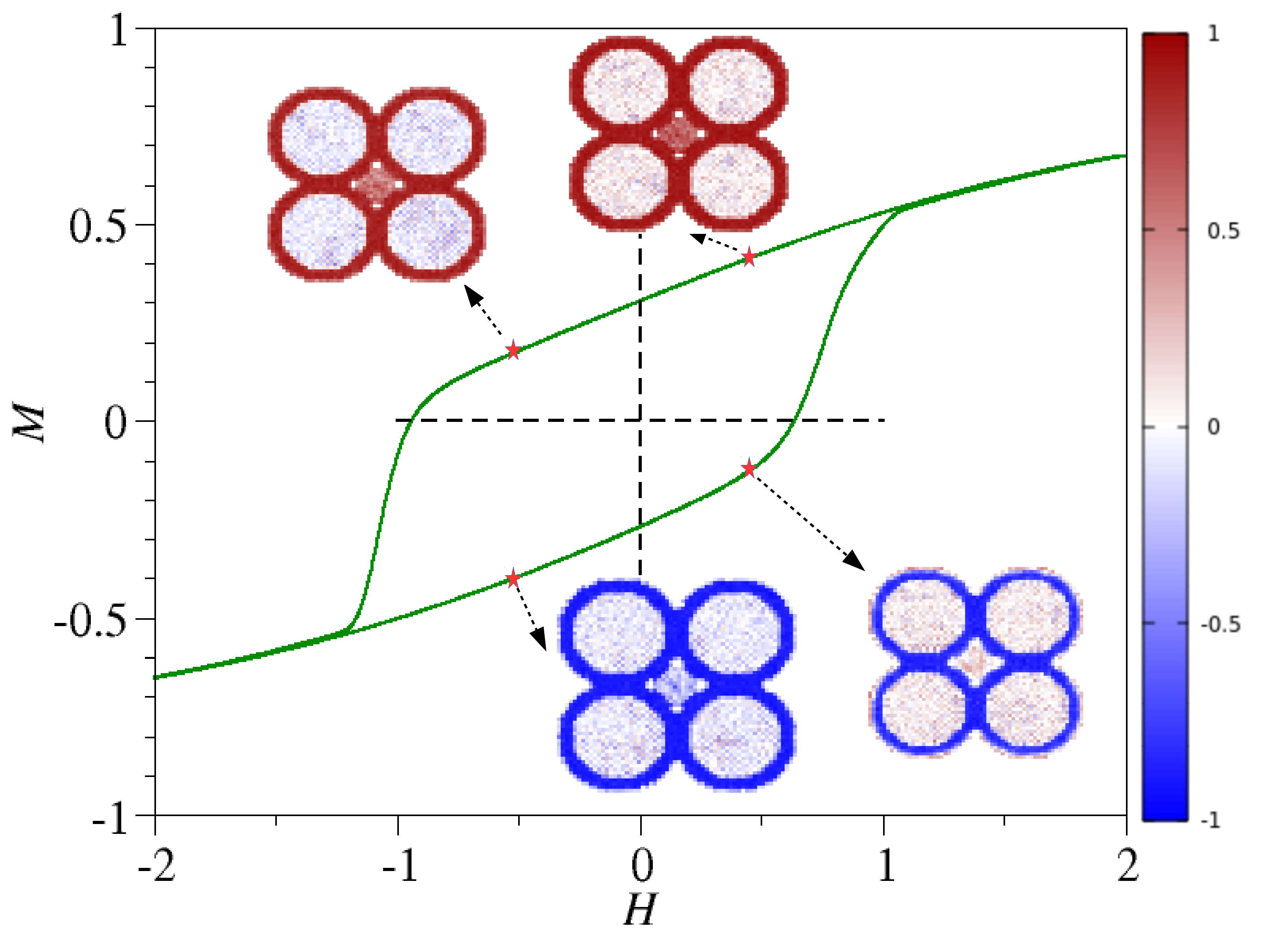}
		\caption{Hysteresis loop for nine nanoparticles clustered as shown in Fig. 1. The  cross-sectional views (at $z=R$) of average local magnetic moment are shown at $H=\pm 0.5$, both in forward and backward directions of the loop. Here the  pinning parameters are $\eta =0.4, r=0.8$ and other parameters are taken same as in Fig. 8.}
\end{figure}

\begin{figure}[ht!]
	\centering
		\includegraphics[scale=0.30]{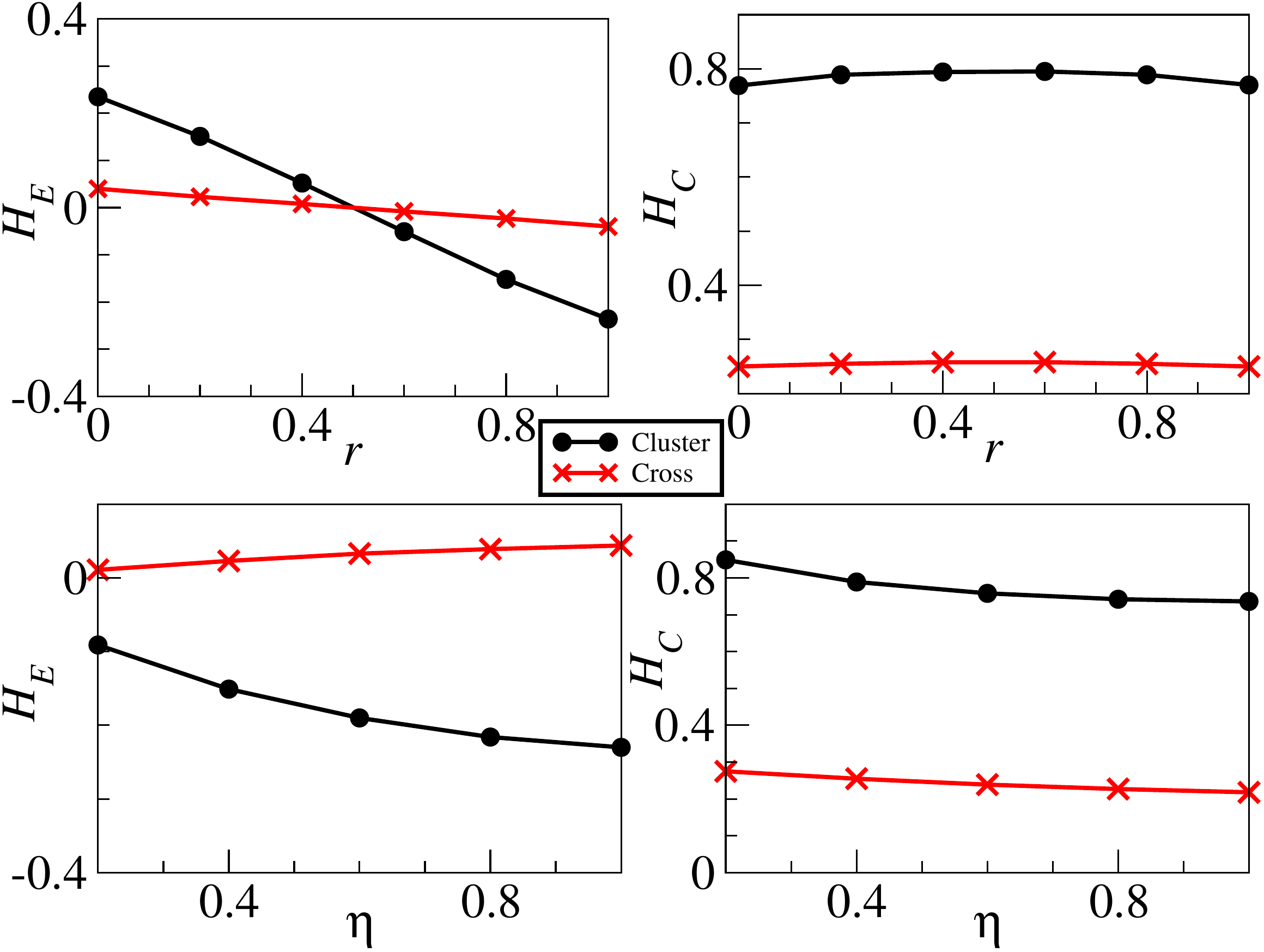}
		\caption{The dependence of $H_E$ and $H_C$:
(top left panel) $H_E$ versus $r$ and (top right panel) $H_C$ versus $r$  for $\eta=0.4$. Symbols $\times$ and $\bullet$ in all the figures represent  respectively the data obtained from Monte Carlo simulations of clustered and crosslinked structures of nine nanoparticles shown in Fig. 1. (bottom left panel) $H_E$ verus $\eta$ and (bottom right panel) $H_C$ versus $\eta$ are shown for  for clustered  nanoparticles  with $r=0.8$ and crosslinked structures with $r=0.2.$}  
		\label{fig:Variation.eps}
\end{figure}

In Figure 12, we show the evolution of the spin structure as a function of applied field and consequent magnetic hysteresis loop for the limited crosslinked structure. We find that the experimentally observed positive exchange bias in such structure could be simulated by assuming 40\% pinning at the surface (i.e., pinning density $\eta$ = 0.4) of which 20\% are oriented preferably along easy-axis (i.e., $r$ = 0.2). The evolution of the color, shown on the particles (Fig. 12), from red to blue and back to red maps the evolution of the moments as a function of applied field within the self-assembled structure. For the structure with massive crosslinking, formed when BBP is used as the dispersing solvent, the experimentally observed negative exchange bias could be simulated (Fig. 13) by assuming the same pinning density yet 80\% spins oriented preferably along easy-axis (i.e., $r$ = 0.8). The other simulation parameters were $R = 16, R_c = 12, K = 3, J_c = -0.5, J_{sh} = 1,  J_{int} = 1$. The simulation further yields dependence of $H_E$ and $H_C$ on $r$ and $\eta$ in these two structures (Fig. 14). In each case, coercivity ($H_C$) decreases but exchange bias ($|H_E|$) increases when $\eta$ is increased. The coercivity ($H_C$) is almost independent of $r$ while $|H_E|$ increases as one moves away from $r = \frac12$. In both the cases, consistent with the experimental observations, it is found that the simulated $|H_E|$ is larger in limited crosslinked structure. The value of coercivity $(H_C)$ in massive crosslinked structure, of course, is greater than that in the limited crosslinked structure. In both the cases, however, number of assembled particles and size of each particles were considered to be identical. This model, therefore, effectively captures the underlying physics of these nanoparticle aggregates. The dispersion of particles within different organic solvents leads to different extent of surface functionalization. The extent of functionalization depends on the particle-liquid interaction potential $\epsilon_{nl}$ and gives rise to the formation of different nanoparticle aggregate structures (Sample 2 and Sample 3) upon drying of the liquid. The surface defects created in the process governs the surface spin pinning potential. The variation in surface spin pinning density for spins oriented or not oriented along the applied field, in turn, yields the change in the magnetic properties of the aggregates including the exchange bias ($H_E$) - its magnitude and sign. The effective magnetic field produced by surface spin pinning is actually responsible for the variation in $H_E$. When superstructure of nanoparticles are produced artificially, their magnetic properties are modified significantly (compared to isolated nanoparticles) due to exchange and dipolar interactions \cite{Givord,Dolci} between particles. But, in the present case, functionalization of particle surface as well as formation of different structures are effected together by the solvents. The particle to particle interaction here is, in fact, mediated by surface functionalization which, in turn, governs the overall surface spin pinning effect across the aggregates of the particles. The influence of particle to particle interaction via surface functionalization is evident in the difference observed in pattern of variation of $H_E$ with pinning density $\eta$ and fraction of surface spins pinned along the direction of applied field $r$ (Fig. 14). Therefore, contrary to the observations made by others \cite{Givord,Dolci}, the magnetic properties, in the present case, are primarily governed by surface spins and not by dipolar or exchange interactions among the particles.   

It is also important to point out here that the exchange coupling interaction $J_{int}$ between the core and shell spins was not found to have any significant impact on the magnitude and sign of the exchange bias. Earlier works \cite{Nogues-1,Nogues-2,Nogues-3} had poined out that positive exchange bias results from competition between antiferromagnetic exchange coupling across ferromagnetic/antiferromagnetic interface (i.e., negative $J_{int}$) and ferromagnetic coupling between applied field and antiferromagnetic moment at the interface. However, as shown in our earlier theoretical work \cite{Sahoo}, neither in two-dimensional nor in three-dimensional spin structure, variation of $J_{int}$ (including switch in $J_{int}$ from positive to negative) has any significant impact on the exchange bias for Ising, XY, and Heisenberg spins. The role of $J_{int}$ in governing the sign and magnitude of exchange bias could be significant in nanoscale core-shell structures where the core is ferromagnetic and the shell is antiferromagnetic and the measurement is carried out, primarily, under field-cooled condition. In fact, switch in $J_{int}$ from positive to negative and competition between exchanged coupled interface spins and antiferromagnetic spin-applied field coupling were shown to reproduce the experimentally observed positive exchange bias in three-dimensional modeling which also takes care of the anisotropy. The surface spin pinning \cite{Berkowitz}, on the other hand, is found to have profound impact in the inverted core-shell structure (where the core is antiferromagnetic and the shell is ferromagnetic) and even when the measurement of magnetic hysteresis loop is done under zero-field-cooled condition \cite{Sahoo}. As shown here, extension of the work even for aggregates of particles emphasizes the role of surface spin pinning with respect to that of exchange interaction $J_{int}$. The spin structure of the nanoscale samples and the measuement protocol used in the present work, therefore, conform to these conditions very well. Our earlier work \cite{Sahoo} has already discussed the role of surface pinning potential, pinning density, and the relative density of $\uparrow$ and $\downarrow$ spins in Ising system. In the present case, the simulation results presented for spherical nanoparticles with Heisenberg spins are found to reproduce the experimental observations quite appropriately. It points out that dispersion of particles within different liquid media for template-free formation of nanoparticle aggregates is actually giving rise to the variation in the surface/interface structures to yield variation in the above-mentioned parameters such as pinning potential, pinning density, and density of pinned spins oriented preferably along the easy-axis. They, in turn, influence the magnetic properties such as coercivity, exchange bias, magnetization siginificantly. Impact of these parameters on the overall multiferroicity (i.e., ferroelectricity, magnetization, and coupling between them) in such functionalized nanoparticles aggregates will be investigated in near future.

\section{Conclusion}

We observed four fold enhancement of coercivity and structure-dependent spontaneous exchange bias in aggregates of BiFeO$_3$ nanoparticles. The aggregation takes place through the drying mediated process where evaporation of the liquid drives the formation of the nanostructures from the nanoparticles suspended in the liquid. Simulation of the process shows that liquid-nanoparticle interaction potential and biasing of evaporation along a certain direction play key role in formation of the nanostructures observed experimentally. Simulation of the surface/interface spin structure and interactions also show that the observed variation in coercivity and exchange bias results from variation in the surface spin pinning potential, pinning density, and density of pinned spins oriented preferably along the easy-axis.  Large and structure-dependent exchange bias could be extremely useful for applications in data storage devices. Since BiFeO$_3$ exhibits electric field driven switching of magnetization, observation of structure-dependent exchange bias field is expected to augment the functionality of the devices by a great extent. Dependence of properties on structures of nanoparticle aggregates offers an additional tool to tune them by designing different structures.\\ 

\noindent $\textbf{SUPPLEMENTARY MATERIAL}$\\

It contains additional TEM and MFM images as well as results of magnetic measurements as a function of time and temperature. The movies showing the simulated evolution of patterns via drying mediated process over a time scale of 20s are also included. \\

\noindent $\textbf{ACKNOWLEDGMENTS}$\\

The author S.G. acknowledges Senior Research Associateship of CSIR, Govt of India during this work. The author A.S. acknowledges DST-INSPIRE fellowship, Govt of India. The author P.K.M. acknowledges support from Science and Engineering Research Board (SERB), India, grant no. TAR/2018/000023 and finally, the author D.B. acknowledges support from CSIR supra-Institutional project `GLASSFIB' (ESC0202).\\

\noindent $\textbf{DATA AVAILABILITY STATEMENT}$\\

The data that supports the finding of this study are available within the article and its supplementary material. The supplementary material including the videos are available from the authors on request.\\

\noindent $\textbf{REFERENCES}$

\end{document}